\documentclass{article}

\usepackage{arxiv}

\usepackage{algorithm}
\usepackage{algpseudocode}
\usepackage[utf8]{inputenc} 
\usepackage[T1]{fontenc}    
\usepackage{hyperref}       
\usepackage{url}            
\usepackage{booktabs}       
\usepackage{amsfonts}       
\usepackage{nicefrac}       
\usepackage{microtype}      
\usepackage{lipsum}		
\usepackage{graphicx}
\usepackage{natbib}
\usepackage{doi}

\usepackage{amsmath}
\usepackage[utf8]{inputenc} 
\usepackage[T1]{fontenc}    
\usepackage{hyperref}       
\usepackage{url}            
\usepackage{booktabs}       
\usepackage{amsfonts}       
\usepackage{nicefrac}       
\usepackage{microtype}      
\usepackage{xcolor}         

\definecolor{peachfuzz}{RGB}{255, 190, 152}

\usepackage{orcidlink}

\usepackage{algorithm}
\usepackage{algpseudocode}
\usepackage{comment}
\usepackage{cleveref}

\usepackage{amssymb}
\usepackage{subcaption}
\usepackage{xspace}

\usepackage{lscape}
\usepackage{graphicx}
\usepackage{pifont}

\usepackage{bbold}
\usepackage{mathtools}
\usepackage{framed}
\usepackage{pifont}
\usepackage{cleveref}
\usepackage{algorithm}
\usepackage{booktabs}
\usepackage{multirow}
\usepackage{hyperref}
\usepackage[utf8]{inputenc}
\usepackage{adjustbox}
\usepackage{subcaption}
\usepackage{url}
\usepackage{soul}
\usepackage{xspace}

\usepackage{multicol}

\usepackage{xcolor}
\usepackage{xargs}
\usepackage[colorinlistoftodos,prependcaption,textsize=small]{todonotes}
\usepackage[framemethod=tikz]{mdframed}

\newcommandx{\unsure}[2][1=]{\todo[linecolor=red,backgroundcolor=red!25,bordercolor=red,caption={},#1]{#2}}
\newcommandx{\change}[2][1=]{\todo[linecolor=blue,backgroundcolor=blue!25,bordercolor=blue,caption={},#1]{#2}}
\newcommandx{\info}[2][1=]{\todo[linecolor=green,backgroundcolor=green!25,bordercolor=green,caption={},#1]{#2}}
\newcommandx{\improvement}[2][1=]{\todo[linecolor=purple,backgroundcolor=purple!25,bordercolor=purple,caption={},#1]{#2}}
\newcommandx{\discussion}[2][1=]{\todo[linecolor=blue,backgroundcolor=yellow!25,bordercolor=yellow,caption={},#1]{#2}}

\newcommandx{\thiswillnotshow}[2][1=]{\todo[disable,#1]{#2}}

\hypersetup{
    unicode=false,          
    pdftoolbar=true,        
    pdfmenubar=true,        
    pdffitwindow=false,     
    pdfstartview={FitH},    
    pdftitle={My title},    
    pdfauthor={Author},     
    pdfsubject={Subject},   
    pdfcreator={Creator},   
    pdfproducer={Producer}, 
    pdfnewwindow=true,      
    colorlinks=true,       
    linkcolor=black,          
    citecolor=black,        
    filecolor=black,      
    urlcolor=black           
}

\newcommand{\spara}[1]{\smallskip\noindent{\bf #1}}

\newcommand{\shiftingmatrix}{\ensuremath{\mathcal{S}^u}\xspace}

\newcommand{\pickingprob}{P\ensuremath{_t(i|u)}\xspace}
\newcommand{\pickingprobfromlist}{P\ensuremath{^L_t(i|u)}\xspace}
\newcommand{\pickingprobfromcatalog}{P\ensuremath{^I_t(i|u)}\xspace}
\newcommand{\recscoreprob}{P\ensuremath{^s_t(i|u)}\xspace}
\newcommand{\inertia}{P\ensuremath{^o(i|u)}\xspace}
\newcommand{\pickeditem}{\ensuremath{i^*}\xspace}
\newcommand{\ctarget}{\ensuremath{\tilde{c}}\xspace}

\newcommand{\squishlist}{
 \begin{list}{$\bullet$}
  {  \setlength{\itemsep}{0pt}
     \setlength{\parsep}{3pt}
     \setlength{\topsep}{3pt}
     \setlength{\partopsep}{0pt}
     \setlength{\leftmargin}{2em}
     \setlength{\labelwidth}{1.5em}
     \setlength{\labelsep}{0.5em}
} }
\newcommand{\squishlisttight}{
 \begin{list}{$\bullet$}
  { \setlength{\itemsep}{0pt}
    \setlength{\parsep}{0pt}
    \setlength{\topsep}{0pt}
    \setlength{\partopsep}{0pt}
    \setlength{\leftmargin}{2em}
    \setlength{\labelwidth}{1.5em}
    \setlength{\labelsep}{0.5em}
} }

\newcommand{\squishdesc}{
 \begin{list}{}
  {  \setlength{\itemsep}{0pt}
     \setlength{\parsep}{3pt}
     \setlength{\topsep}{3pt}
     \setlength{\partopsep}{0pt}
     \setlength{\leftmargin}{1em}
     \setlength{\labelwidth}{1.5em}
     \setlength{\labelsep}{0.5em}
} }

\newcommand{\squishend}{
  \end{list}
}
\title{Algorithmic Drift: A Simulation Framework to Study the Effects of Recommender Systems on User Preferences}    
\author{%
  Erica Coppolillo\thanks{E. Coppolillo and S. Mungari equally contributed to the paper.} \\
  University of Calabria\\
  ICAR-CNR \\
  \texttt{erica.coppolillo@unical.it} \\
  \And
  Simone Mungari \\
  University of Calabria\\
  ICAR-CNR\\
  Revelis s.r.l.\\
  \texttt{simone.mungari@unical.it} \\
  \AND
  Ettore Ritacco \\
  University of Udine\\
  \texttt{ettore.ritacco@uniud.it} \\
   \And
  Francesco Fabbri \\
  Spotify\\
  \texttt{francescof@spotify.com} \\
   \And
  Marco Minici \\
  University of Pisa\\
  ICAR-CNR\\
  \texttt{marco.minici@icar.cnr.it} \\
   \AND
  Francesco Bonchi \\
  CENTAI\\
  \texttt{francesco.bonchi@centai.eu} \\
  \And
  Giuseppe Manco \\
  ICAR-CNR\\
  \texttt{giuseppe.manco@icar.cnr.it} \\
}

\renewcommand{\shorttitle}{Algorithmic Drift}

\begin{document}

\maketitle

\begin{abstract}
Digital platforms such as social media and e-commerce websites adopt Recommender Systems to provide value to the user.
However, the social consequences deriving from their adoption are still unclear.
Many scholars argue that recommenders may lead to detrimental effects, such as bias-amplification deriving from the feedback loop between algorithmic suggestions and users’ choices. Nonetheless, the extent to which recommenders influence changes in users leaning remains uncertain. In this context, it is important to provide a controlled environment for evaluating the recommendation algorithm before deployment. To address this, we propose a stochastic simulation framework that mimics user-recommender system interactions in a long-term scenario. In particular, we simulate the user choices by formalizing a \textit{user model}, which comprises behavioral aspects, such as the user \textit{resistance} towards the recommendation algorithm and their \textit{inertia} in relying on the received suggestions. Additionally, we introduce two novel metrics for quantifying the algorithm's impact on user preferences, specifically in terms of drift over time. We conduct an extensive evaluation on multiple synthetic datasets, aiming at testing the robustness of our framework when considering different scenarios and hyper-parameters setting. The experimental results prove that the proposed methodology is effective in detecting and quantifying the drift over the users preferences by means of the simulation. All the code and data used to perform the experiments are publicly available\footnote{\url{https://anonymous.4open.science/r/AlgorithmicDrift-D553}}.
\end{abstract}



\section{Introduction}
\label{sec:intro}

Nowadays, people adopt social media as a fundamental medium to share and consume information. 
With a growing amount of available content, 
recommender systems represent a viable way to help users navigate large volumes of information by suggesting content that the user may like.
As a downside, recommendation algorithms have also been blamed for detrimental effects such as echo chambers~\citep{pariser2011filter} and opinion polarization~\citep{cho2020search}, 
which in turn may lead to pernicious phenomena such as misinformation spreading~\citep{del2016spreading}, fragmentation~\citep{sunstein2018republic}, and radicalization~\citep{sunstein1999law}.
As a natural consequence, 
measuring the potential side-effects of recommender systems in the emergence of these issues is attracting much interest, 
with scholars proposing data-driven analysis~\citep{bakshy2015exposure, de2021no, cinelli2021echo}, 
model-based methods~\citep{minici2022echochamber}, 
and simulation studies~\citep{fabbri2020effect, santos2021link, cinus2022effect, tommasel2022recommender}.

Despite this growing interest, the challenge is still open in the current literature in analyzing and measuring how 
the algorithm effects on users may evolve over time. Although several models have been proposed~\citep{chaneyHowAlgorithmicConfounding2018, siren, yao2021measuringrecommendereffectssimulated,relevancemeetsdiversity}, these studies focus on specific topics, such as behavioral homogeneity or items diversity. Indeed, there is still lack of a unified and generalizable methodology which allows to track users preferences' evolution over time, and that is orthogonal to the kind of drift to quantify. Further, most of the existing simulation environments follow a simplistic behavioral model, providing limited flexibility in representing different behavioral patterns.


To fill this gap, in this paper, we propose a stochastic model for characterizing time-based interactions between users and recommendation algorithms in operational environments.
In our scenario, users interact with various items on a platform, such as media content, news articles, or videos. The platform collects data on user preferences and utilizes this information to generate personalized recommendations. These recommendations are typically provided by ranking items based on their predicted relevance to the user~\citep{trust_bias}.


An important aspect of our formulation consists in accounting for the "feedback loop" effect described in the literature~\citep{chaneyHowAlgorithmicConfounding2018}, where user choices are influenced by the recommendations provided, and the algorithm relies on the user's past interactions rather than aligning with their true interests. Therefore, to accurately simulate user choices, it is essential to characterize user behavior by considering several factors, such as their potential \textit{resistance} to recommendations (i.e., the autonomous selection of an item from the entire catalog) and their \textit{inertia} in following the provided suggestions.
Starting from this, the research questions we aim to address are the following: \textit{Can we quantify the effect of the recommendation algorithm in altering user preferences over the long term? How do different behavioral users pattern affect the influence of the recommender?}

To answer these questions, we propose a controlled simulated environment where mimicking user-recommender system interactions, along with two novel metrics to evaluate whether and to what extent the recommender system contributes to drifting users' initial preferences.
Following from these assumptions, we introduce the concept of ``\emph{algorithmic drift}'', in order to characterize how the recommendation algorithm contributes in changing  user leanings. In practice, the simulation model starts from an initial group of heterogeneous user preferences, from which the recommender system induces initial transition probabilities between item categories.  
Notably, besides (blindly) following the provided recommendations, our model allows users to either completely ignore the suggestions and pick an item autonomously from the catalog (\textit{resistance}), or still examine the recommendation list but choose the item still relying on their own preferences (\textit{inertia}). Further, the user's choice can be influenced by exogenous hence unpredictable factors (\textit{randomness}), that can lead to spurious interactions.
By supporting different behavioral patterns in the users choices, we provide a comprehensive framework that allows for measuring the effects of the recommendation algorithms under different conditions. Notably, the metrics we propose enable such measurements, by tracking and quantifying changes according to predefined criteria. 

As a paradigmatic example of such criteria, in our experiments we assume that the items available on the platform can be categorized as either \emph{harmful} or \emph{neutral}. Simultaneously, users are classified into three categories based on their interactions with these items: \textit{non-radicalized}, \textit{semi-radicalized}, or \textit{radicalized}. This classification is determined by the proportion of harmful interactions they have exhibited. Further, we focus on a collaborative filtering-based scenario, where we consider the implicit connections that can be induced by the collaborative nature of the algorithm, hence exploiting the bridging nature of some users to propagate influences on the adoptions. As a result, the proposed framework allows us to quantify the changes in user leanings.

Our contributions can be summarized as follows:

\begin{itemize}
    \item We define the concept of \emph{algorithmic drift} and introduce two metrics for evaluating the impact of recommender systems on user behavior changes in the long term.
    \item We model a novel simulation framework and standard methodology to study the effect of any recommendation algorithm in a completely model-driven setting.
    \item We conduct an extensive analysis to assess the effectiveness of our methodology and its robustness across different scenarios.
\end{itemize}

The rest of the paper is structured as follows. Section~\ref{sec:related} illustrates the current literature in the context of recommendation. In Section~\ref{sec:model}, we formally describe the proposed simulation framework. Section~\ref{sec:experiments} presents the results of the experimental evaluation. Finally, Section~\ref{sec:conclusions} concludes the paper, depicting some pointers for future work.

\begin{figure}[!ht]
    \centering
    \includegraphics[width=\columnwidth]{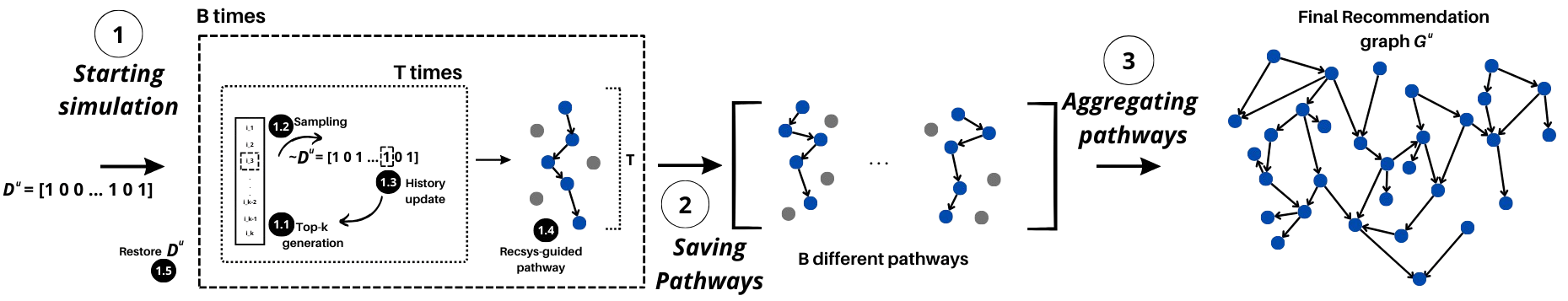}
    \caption{{Framework overview. Considering the initial interactions of each user $u$, we start the simulation (1) by invoking the trained recsys model B times. In each of these iterations, we get the top-k recommended items (1.1) and we sample an item (1.2), thus updating the history of $u$ (1.3). We repeat the process T times, in order to build a tree whose nodes are the items sampled (1.4), of depth T. Intuitively, each tree represents a pathway of the user $u$ guided by the recsys. After each iteration, we restore the initial user history (1.5), and the process restarts. At the end of the B iterations, we obtain $T$ different trees, representing each pathway (2). Hence, we aggregate all the trees (3) and we obtain the final recommendation graph $G^u$.
    }}
    \label{fig:toyexample}
\end{figure}

\section{Related work}
\label{sec:related}

The contribution of this work spans on a variety of research fields, whose literature is explored as follows.



\spara{Simulation-based studies.}
We can observe a growing research line studying the long-term effects of recommender systems through simulation models.
\citet{PrCP} propose a simulation framework for studying polarization phenomena in a context in which users are binarized into two homogeneous groups.

\citet{chaneyHowAlgorithmicConfounding2018} claim that recommenders try to homogenize user behavior without increasing the utility of their suggestions. They argue that the most recent recommendation systems are actually trained over preferences that are already biased by previous recommendations, thus triggering a feedback loop where a suggestion algorithm tries to fit the recommendation history instead of matching the real interests of the users.

\citet{siren} propose a simulation framework for studying the impact of recommender systems in drifting users towards different content topics, thus mainly focusing on items diversity and users serendipity. Similarly, ~\citet{yao2021measuringrecommendereffectssimulated} depicts a simulated environment to assess the actual contribution of recommenders in modifying users habits, particularly focusing on the phenomenon of popularity bias. 

\citet{fabbri2022exposure, cinus2022effect, santos2021link} show how people-recommenders can exacerbate social media critical issues, such as polarization, misinformation, and pre-existing inequalities in user communities. The authors define a simulation model that highlights the following recommender systems effects: (i) the growth of exposure inequalities at the individual level, strengthening the “rich-get-richer” effect \cite{fabbri2022exposure}, and (ii) the formation of polarized communities and echo chambers within homophilic groups \cite{cinus2022effect, santos2021link}.
In a similar vein, \cite{DBLP:journals/corr/abs-2102-00099} devised a model for simulating changes in users' opinions in social networks, showing potential detrimental effects such as polarization and echo chambers formations.

\citet{tommasel2022recommender} focus on the misinformation spread in social networks due to the recommender system capability of influencing the network topology. First, they simulate changes in user misinformation-spreading behavior and define counterfactual scenarios.
Then, they simulate an opinion dynamic model over the derived network to estimate how recommendations affect the influence of users spreading misinformation.

We would like to emphasize how our work draws inspiration from the discussed literature but substantially differs from it. Indeed, we are not interested in explaining or assessing detrimental phenomena induced by recommender systems, but we aim at providing a controlled environment where the algorithm can be tested before deployment, by simulating its interaction with the users in the long term.

\spara{Data-driven studies.}
\citet{ribeiro2020auditing} and \citet{haroon2022youtube} show that YouTube recommendations lead to the genesis of ``\emph{rabbit holes}'', i.e., a journey toward increasingly radicalized contents that can exaggerate user bias, belief, and opinion.
\citet{ribeiro2020auditing} analyze videos on channels belonging to four political leanings. The authors observe that some center channels serve as gateways to push users to far-right ideology, thus showing evidence of radicalization phenomena where users migrate from milder to extreme ideologies. On the other hand, \citet{haroon2022youtube}'s analysis is conducted by exploiting sock puppets (i.e., brand new users with ad-hoc history) which are actively monitored while using YouTube, in order to discover and quantify the emergence of ideological biases.
Similarly, \citet{phadke2022pathways} provide empirical modeling for analyzing the radicalization phenomenon over conspiracy theory discussions in Reddit. As a result, users on increasing conspiracy engagement pathways progress successively through various radicalization stages.

Once again, our proposal shows strong differences with respect to this literature. One of our main contributions is the definition of general quantitative indexes to measure users' opinion drift induced by several CF recommender systems.

\spara{Model-based approaches.}
User behavior deviation from natural evolution has been widely studied in the literature, still remaining an open challenge.
\citet{10.1145/3289600.3291002} define a strategy to build \emph{antidote data} to feed models aiming at promoting the fairness.
To this purpose, authors define a metric to measure user homogeneity as the average variance of the preferences on items. \citet{10.1145/3460231.3478851} model social graphs and user belief as time-dependent processes, where ideological changes and network topology co-evolution in time is monitored.
Their idea is to start from a real social network (e.g., Twitter) and apply changes guided by people-recommenders: thus, drawing inspiration from~\cite{Degroot74, doi:10.1073/pnas.1217220110}, they apply the Duclos-Esteban-Ray polarization measure~\cite{10.2307/3598766} to estimate how recommendation-driven conformities would form. \citet{minici2022echochamber} propose a probabilistic generative model that discovers echo-chambers in a social network by introducing a Generalized Expectation Maximization algorithm, that clusters users according to how information spread among them. \citet{chang2022recommend} define a counterfactual model called ``organic model'', which is used to compare user outcomes influenced by a recommender against the outcomes under the natural evolution of the users' choices. 
The main finding of this work states that the recommender and organic model dramatically differ from each other, thus highlighting the influence of the recommendation.

Differently, our proposal is to formalize a standard methodology, to measure deviations (triggered by any target recommender system) from user normal evolution, that exhaustively navigates the user choice space, instead of considering only one pathway that a user can choose.

\section{Simulation Framework}
\label{sec:model}
The core of our approach consists in devising the simulation framework and a set of metrics for measuring the potential algorithmic drift that a target recommendation algorithm may induce. In the following, we describe such contributions. 

\subsection{Preliminaries}\label{sec:preliminaries}

Let $U$ (resp. $I$) be the set of users (resp. items). We consider an implicit-feedback matrix $\mathcal{D} \in \{0,1\}^{|U|\times|I|}$ of user-item interactions.
$\mathcal{D}_{ui}$ is equal to 1 if user $u$ has interacted with item $i$, and $0$ otherwise. With an abuse of notation, we denote by $\mathcal{D}_{u}\subseteq I$ the set of all items $i$ such that $\mathcal{D}_{ui}=1$. Also, $(u,i) \in \mathcal{D}$ whenever $\mathcal{D}_{ui}=1$.

We assume a labeling function $\ell : I \rightarrow \{{c_1}, \ldots, {c_N} \}$ that tags an item $i$ 
as belonging to a category among $c_1, \ldots, c_N$, thus 
partitioning $I$ into $N$ disjoint subsets $I_{j}=\{i \in 
I| \ell(i)={c_j}\}, j=1, \ldots, N$.

We adopt a recommender system $\mathcal{R_{\theta}}: U \times I \rightarrow [0,1]$, parameterized by $\theta$ and previously trained on $\mathcal{D}$. Given an unseen user-item interaction $(u,i)$, i.e.,  $D_{ui} = 0$, $\mathcal{R}_{\theta}(u,i)$ estimates how likely the user $u$ interacts with an item $i$ once it is exposed to it. Such a likelihood can be exploited to build a ranked recommendation list for a user $u$, formed by unseen items and sorted by decreasing likelihood. 


\subsection{Organic Model}\label{sec:organic_model}
To assess the effect of recommender algorithms in terms of preference alteration, 
it is crucial to have a counterfactual based on an organic model --- i.e., assuming users and items interact without any $\mathcal{R}_{\theta}$.
For this purpose, we adopt the organic model proposed by \citet{chang2022recommend}.
Assuming to know user and item features, $\rho_u$ and $\alpha_i$, respectively, preferences are generated according to the measure $\|\rho_u - \alpha_i\|$. 
However, users don't have full knowledge of the actual features $\alpha_i$ for each item $i$ and can only access samples through a Normal distribution $\hat{\alpha_i}\sim \textrm{Normal}(\alpha_i, \Sigma)$, 
with $\Sigma = 0.5\cdot\Sigma_{\text{item}}$ and $\Sigma_{\text{item}} = \alpha^T\alpha$ being the empirical covariance of the item features. 
\citet{chang2022recommend} also propose to use an interpolation between the sample $\hat{\alpha_i}$ and
the mean of all the samples: 
\begin{equation}
    \hat{\alpha_i}^S = \xi \hat{\alpha_i} + (1-\xi)\left(\frac{1}{|I|}\sum_{j\in I} \hat{\alpha_j}\right)
\end{equation}
We use the shrunken estimate using a value of $\xi=0.4$ since it was empirically determined to be optimal.
However, similar results are obtained by putting the value of $\xi=0$, thus disregarding a global effect in the noisy estimate. 


\subsection{User Behaviour}\label{sec:user_behaviour}
Our simulation model mimics how user preferences and recommender system co-evolve in a long-term scenario. In each step, the user selects an item to interact with, by either resorting to the algorithm recommendations, or by autonomously examining the whole catalog. 
Consequently, the algorithm adapts the next recommendations relying on these new interactions: this tight relationship between users and recommender system models the well-known concept of feedback loop~\citep{chaneyHowAlgorithmicConfounding2018}.
Further, our framework aims at simulating the user navigation across the whole choice space, instead of a single pathway. For this reason, the procedure consists of $T$ iterative trials, each of which is repeated for $B$ independent rounds. We impose that, in each of the $B$ rounds, a user cannot interact with an item more the once. 
Within the simulation, for each user $u$, we instantiate an item-item matrix $S^u \in \mathbb{N}^{|I|\times|I|}$. Intuitively, $S^u_{i,j}$ counts how many times $u$ shifted from item $i$ to item $j$ in two consecutive rounds of the simulation, as explained in the following.


At each step $t$, The recommender system provides a list $L_t\subset I$ of $k$ items to the user. The latter behaves according to the following process: if their \textit{resistance} to the suggestions of the recommender system (quantified as $\gamma$) is high, the user may (stochastically) select an item from the whole catalog, by either relying on their own preferences, or  picking it randomly. The random factor $\eta$ (which we assume to be very low in practice) models the influence of exogenous hence uncontrollable factors that can guide the user's choice in that round. For instance, the user may select a spurious item under the influence of a friend's suggestion, by relying on advertisements, or even by mistake. 

Alternatively, if the user is not highly resistant to the algorithm, they may examine the recommendation list and pick an item. This can occur by either totally relying on the system (high \textit{inertia}), or under the influence of their own interests (low \textit{inertia}).
Here, the core idea is that the user's choice may be influenced by its own preferences and beliefs, or conversely, they may completely rely on the recommender system, thus resembling the trust bias phenomenon \citep{trust_bias}.

Therefore, the probability for user $u$ to select item $i\in I$ at step $t$ is given by:

\begin{equation}
\centering
    \pickingprob = \gamma \cdot \pickingprobfromcatalog + (1 - \gamma) \cdot \pickingprobfromlist
\label{eq:picking_prob}
\end{equation}

where
\begin{align}
    \pickingprobfromcatalog  & = \eta \cdot \frac{1}{|I|} + (1 - \eta) \cdot \inertia 
\label{eq:picking_prob_from_catalog} \\
\pickingprobfromlist & = \left\{\begin{array}{lr}
   \delta \cdot \recscoreprob + (1-\delta) \cdot \inertia &  \mbox{if $i \in L_t$}\\
    0 & \mbox{otherwise}
\end{array}\right.
\label{eq:picking_prob_from_list}
\end{align}



\begin{algorithm}
\caption{Users-Recsys Interaction Process}\label{alg:interaction_process}
\small
\begin{algorithmic}[1]
\State \textbf{Input:} Number of rounds $T$, user set $U$, item set $I$, history dataset $\mathcal{D}$, user matrices $\mathcal{S}^{u_1}, \ldots, \mathcal{S}^{u_{|U|}}$,
recommender system $\mathcal{R}_{\theta}$, resistance factor $\gamma$, random factor $\eta$, inertia factor $\delta$
\State \textbf{Output:} ${G}^{u_1}, \ldots, {G}^{u_{|U|}}$
\For{$b \in \{1, \ldots, B\}$} \Comment{\textit{B} independent trials}
\State $\mathcal{\hat{D}} \gets \mathcal{D}$ \Comment{Revert the original history dataset}
\For{$t \in \{1, \ldots, T\}$}
    \For{$u \in U$} 
    \State $\hat{I}^u \gets \emptyset$
    
    \State $j \gets \textrm{None}$ \Comment{Initialize previously selected item}
    \If{$\textrm{Bernoulli}(\gamma)$} \Comment{Resistance factor $\gamma$}
        \If{$\textrm{Bernoulli}(\eta)$} \Comment{Random factor $\eta$}
            \State Pick $\pickeditem$ uniformly from $I$
        \Else
            \State Pick $\pickeditem$ with probability $\inertia$, $i \in I$
            \Comment{Sampling from catalog based on preferences}
        \EndIf
    \Else
        \State $L = [i_1, i_2, \dots, i_k] \gets \mathcal{R}_{\theta}(\mathcal{\hat{D}})$ \Comment{Top-\textit{k} recommendations}
        \If{$\textrm{Bernoulli}(\delta)$}
        \Comment{Inertia factor $\delta$}
        \State Pick $\pickeditem$ with probability $\recscoreprob$, $i \in L$
        \Comment{Sampling from list based on recommender scores}
        \Else 
        \State Pick $\pickeditem$ with probability $\inertia$, $i \in L$
        \Comment{Sampling from list based on preferences}
        \EndIf
    \EndIf
    \State $\mathcal{\hat{D}}_u \gets \mathcal{\hat{D}}_u \cup \{\pickeditem\}$ \Comment{Updating the history dataset}
    \If{$j$ is not None}
        \State $\mathcal{S}^u_{j, \pickeditem} \gets \mathcal{S}^u_{j, \pickeditem} + 1$    
        \Comment{Updating the user matrix}
    \EndIf
    \State ${\hat{I}}_u \gets \hat{I}_u \cup \{\pickeditem\}$ 
    \State $j \gets \pickeditem$
\EndFor
\EndFor
\EndFor
\State $\mathcal{\hat{S}}^{u_1}, \ldots, \mathcal{\hat{S}}^{u_{|U|}}  \gets$ Normalize($\mathcal{{S}}^{u_1}, \ldots, \mathcal{{S}}^{u_{|U|}}$) \Comment{Computing Row-wise Normalization of user matrices}
\State ${G}^{u_1}, \ldots, {G}^{u_{|U|}} \gets (\hat{I}^u_1, \mathcal{{S}}^{u_1}), \ldots, (\hat{I}^{u_{|U|}}, \mathcal{{S}}^{u_{|U|}})$ 
\Comment{Building users probabilistic graphs}
\end{algorithmic}
\end{algorithm}

%

Here, $\recscoreprob$ represents the probability of selecting the item based on the score provided by the recommendation 
algorithm, while $\inertia$ is the probability of picking it by following the natural preferences of the user. We define them as:


\begin{align}
    \recscoreprob &= \frac{\mathcal{R}_{\theta}(i)}{\sum_{j\in L_t} \mathcal{R}_{\theta}(j)}
\label{eq:recscoreprob} \\
    \inertia &= \frac{\|\rho_u - \hat{\alpha_i}^S\| - \min_{j\in L_t}(\|\rho_u - \hat{\alpha_j}^S\|)}{\max_{j\in L_t}(\|\rho_u - \hat{\alpha_j}^S\|) - \min_{j\in L_t}(\|\rho_u - \hat{\alpha_j}^S\|)}
\label{eq:inertia}
\end{align}


The linear combination of these two components is weighted in Equation~\ref{eq:picking_prob} using the \textit{inertia} parameter $\delta \in [0, 1]$. Intuitively, if $\delta=1$, the user blindly follows the recommendations provided at each round, resembling the trust bias phenomenon \citep{trust_bias}; conversely, if $\delta = 0$, the user selects the item based on their preferences only, i.e., by following the organic model; finally, if $0 < \delta < 1$, the choice is conditioned by both the user's interests and the recommendation score.
We denote the selected item as \pickeditem.
Within the modeled scenario, at each round, the choice of \pickeditem is tracked within the dataset $\mathcal{D}_u$, which is used by the recommender algorithm to provide new suggestions. This also results in increasing by 1 the value $S^u_{j, \pickeditem}$, where $j$ represents the item that $u$ selected in the previous round.

Notably, the matrix \shiftingmatrix represents a blueprint of how the preferences of a user $u$ co-evolve together with a recommender system $\mathcal{R}_\theta$ 
from a long-term perspective.
Starting from \shiftingmatrix, in fact, we are able to generate a probabilistic graph $G^u=(\hat{I}^u, \hat{\mathcal{S}}^u)$, where $\hat{I}^u \subset I$ is the subset of items for which $u$ had at least one interaction during the simulation, 
and $\hat{\mathcal{S}}^u$ being the row-normalized stochastic shifting matrix. 
Note that $G^u$ captures the effects of the recommender model $\mathcal{R}_\theta$ in the long term, as it exhibits transition probabilities and hence ultimately the likelihood  of users drifting their leanings. 
The final output of our simulation process is hence the probabilistic graph $G^u$, retrieved for each user $u$.
We summarize the whole procedure in Algorithm~\ref{alg:interaction_process}
and illustrate the overall flow in Figure~\ref{fig:toyexample}. In our experiments, we use $B = 50$ and $T = 100$.

As a final note, we model the user-algorithm feedback loop by feeding $\mathcal{R}_\theta$ with the updated matrix $\mathcal{\hat{D}}$ at each step $t$ (line 16 of Algorithm~\ref{alg:interaction_process}). Despite other modeling choices can be considered (e.g., retraining the recommender at each step), we here do not investigate the impact of such alternative implementations, this being orthogonal to our study and beyond the scope of this paper.


\subsection{Evaluation}
Through our simulation framework, we are interested in studying the way user preferences evolve in the long term. 
We refer to this phenomenon as "algorithmic drift", and we further introduce two novel metrics in order to quantify it.

\spara{Algorithmic Drift Score}. As aforesaid, it can be expressed as the tendency of the recommendation algorithm to alter user preferences after multiple interactions.
In other words, assuming that content in a platform can be tagged as belonging to a given category, the drift represents the deviation of the initial users preferences towards other kinds of content, after interacting with the recommendation algorithm ($\mathcal{R}_\theta$).

To quantify the extent of this phenomenon induced by the recommender system, we define the \textit{Algorithmic Drift Score} (\textit{ADS)} over the probabilistic graph $G^u$ of a user $u$.
Such graph-based metric is an adaptation of the Random Walk Controversy Score (RWC), proposed by~\citet{garimella2017quantifying} to quantify the radicalization of user opinions in an online social network.
In our case, 
this metric describes whether the user is more inclined to encounter and remain into pathways of items belonging to a given target category \ctarget ($i \in I_{\ctarget}$), starting from items belonging to a different category ($i \in I_j, j \neq \textrm{target}$).

Let us consider a user $u$, and its probabilistic graph $G^u$. The measure of algorithmic drift $\textit{ADS}(G^u)\in[-N+1, 1]$ can be hence defined as:
\begin{equation}
 \textit{ADS}(G^u) = \prod_{i=1}^N  Pr(I_{\ctarget} | I_{i}) -  \sum_{j=1 \atop j \neq \ctarget}^N \prod_{k=1}^{N} Pr(I_{j} | I_{k}) 
\label{eq:algorithmic_drift}
\end{equation}

where $Pr(X|Y)$ is the probability that a random walk ends on an element belonging to the set $X$, 
given that the walk starts on an element of the set $Y$. 
Intuitively, the larger is the value of $\textit{ADS}(G^u)$, the more likely the user $u$ will consume content belonging to the target category. Vice-versa, the lower value, the higher the probability that user $u$ follows pathways of content from different categories.

\spara{Delta Target Consumption.} To quantify the change in user consumption of a target category before and after interacting with the recommendations, we introduce the \emph{Delta Target Consumption} (\textit{DTC}) rate. 
Let us consider a user $u$, their historical set of interactions $\mathcal{D}_u$, and the set of simulated interactions $\hat{I}_u$. Then, 
\begin{equation} \label{eq:delta_harmful_consumption}
    \textit{\textit{DTC}}(u) = \frac{|I_{\ctarget}\cap (\hat{I}_u \cup \mathcal{D}_u)|}{|\hat{I}_u \cup \mathcal{D}_u|} - 
    \frac{|I_{\ctarget} \cap \mathcal{D}_{u}|}{|\mathcal{D}_{u}|}
\end{equation}
representing the variation of the relative frequency of harmful items after the interaction summarized in $\hat{I}_u$.
The larger the value of $\textit{DTC}(u)$, the more the recommender $\mathcal{R}_\theta$ is shifting the user's choices towards items belonging to the target category \ctarget ($i \in I_{\ctarget}$) with the respect to their initial preferences. 

Intuitively, a positive value for $\textit{ADS}(G^u)$ and/or $\textit{DTC}(u)$ implies that $u$'s preferences have been driven toward more content tagged as \ctarget; vice versa, if $\textit{ADS}(G^u)$ and/or $\textit{DTC}(u) < 0$, $u$ has interacted with less items in $I_{\ctarget}$. Finally, a value equal to $0$ means that the preferences of $u$ have not been altered in the long term with respect to the target category.

\section{Experiments}
\label{sec:experiments}
In this section, we study how the simulation framework and the proposed metrics act when dealing with different behavioral users patterns. Specifically, we aim at empirically addressing the following research questions:
\begin{itemize}
    \item[\textbf{RQ1.}] Can the proposed methodology (simulation framework and related metrics) effectively quantify drifts in users' leaning due to recommendation algorithms? 
    \item[\textbf{RQ2.}] How robust is such an approach to different behavioral users patterns, i.e., resistance, inertia, and randomness?
\end{itemize}

\spara{Motivating Use Case}.\label{sec:use_case}
For the experimental evaluation of our methodology, as a practical use case, we devise a scenario where items are categorized as either \textit{harmful} or \textit{neutral}, i.e., $i \in \{I_h, I_n\}$, and users are divided into three sub-populations: \textit{non-radicalized}, \textit{semi-radicalized}, and \textit{radicalized}, according to the percentage of harmful content they exhibit.
Items are here intended as \textit{harmful} in a broad sense, without referring to any specific semantic field. In other words, they can spread to whatsoever kind of content (noxious, explicit, inappropriate), e.g., violence, misinformation, pornography, and hate speech. In this respect, radicalized users can also be considered as prone to a high consumption rate of harmful content. 

Therefore, in this setting, the objective is evaluating the algorithmic drift induced over users w.r.t. the \textit{harmful} category, i.e., quantifying the radicalizing pathways encountered by non-radicalized users, after interacting with the algorithm in the long-term. In order to do this, Equation~\ref{eq:algorithmic_drift} can be easily adapted by fixing the target category as $h$, as follows:
\begin{equation} 
    \textit{ADS}(G^u) = Pr(I_h | I_h) \cdot Pr(I_h | I_n) - Pr(I_n | I_n) \cdot Pr(I_n | I_h)
\label{eq:radicalization_measure_readapted}
\end{equation} 

Notably, the ADS score can now assume values in the range [-1, 1], these being the corresponding preference poles of non-radicalized and radicalized users, respectively. Therefore, intuitively, the higher the drift effect, the more the value will shift toward the opposite pole. In other words, in presence of significant drift, the ADS computed over a non-radicalized user will approach 1, and vice-versa (the ADS computed over a radicalized user will approach -1).

We can similarly tailor the definition of \textit{DTC} in Equation~\ref{eq:delta_harmful_consumption} (target = $h$) in order to trace the variation of user consumption in terms of harmful content. Intuitively, the higher the rate, the more harmful content the user will consume in the long term.

\subsection{Data Generation}\label{sec:data_generation}

\begin{figure}[th!]
     \centering
     \begin{subfigure}[b]{0.32\textwidth}
         \centering
         \includegraphics[width=\textwidth]{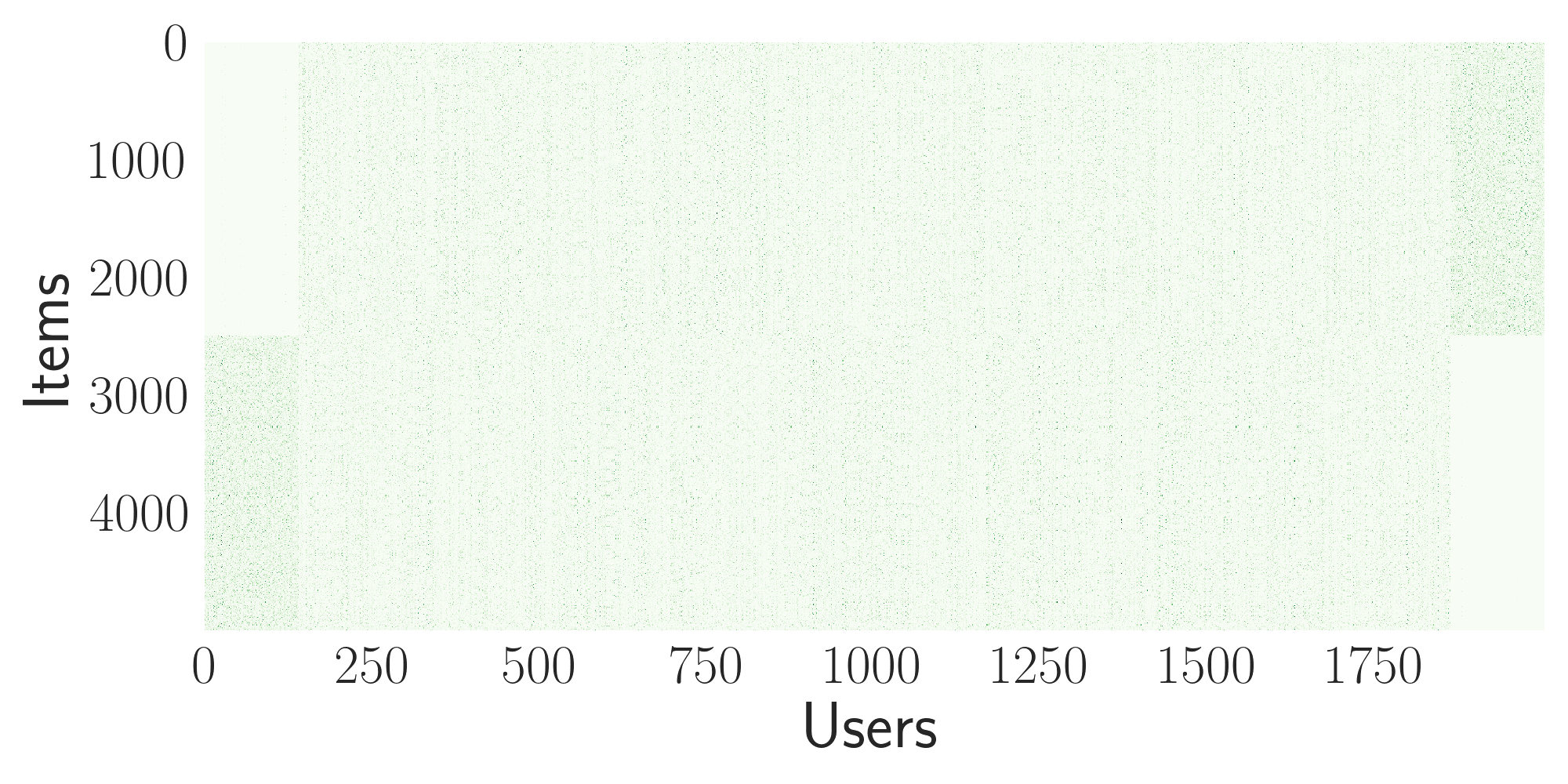}
         \caption{5\%, 90\%, 5\%}
     \end{subfigure}
    \hfill
     \begin{subfigure}[b]{0.32\textwidth}
         \centering
         \includegraphics[width=\textwidth]{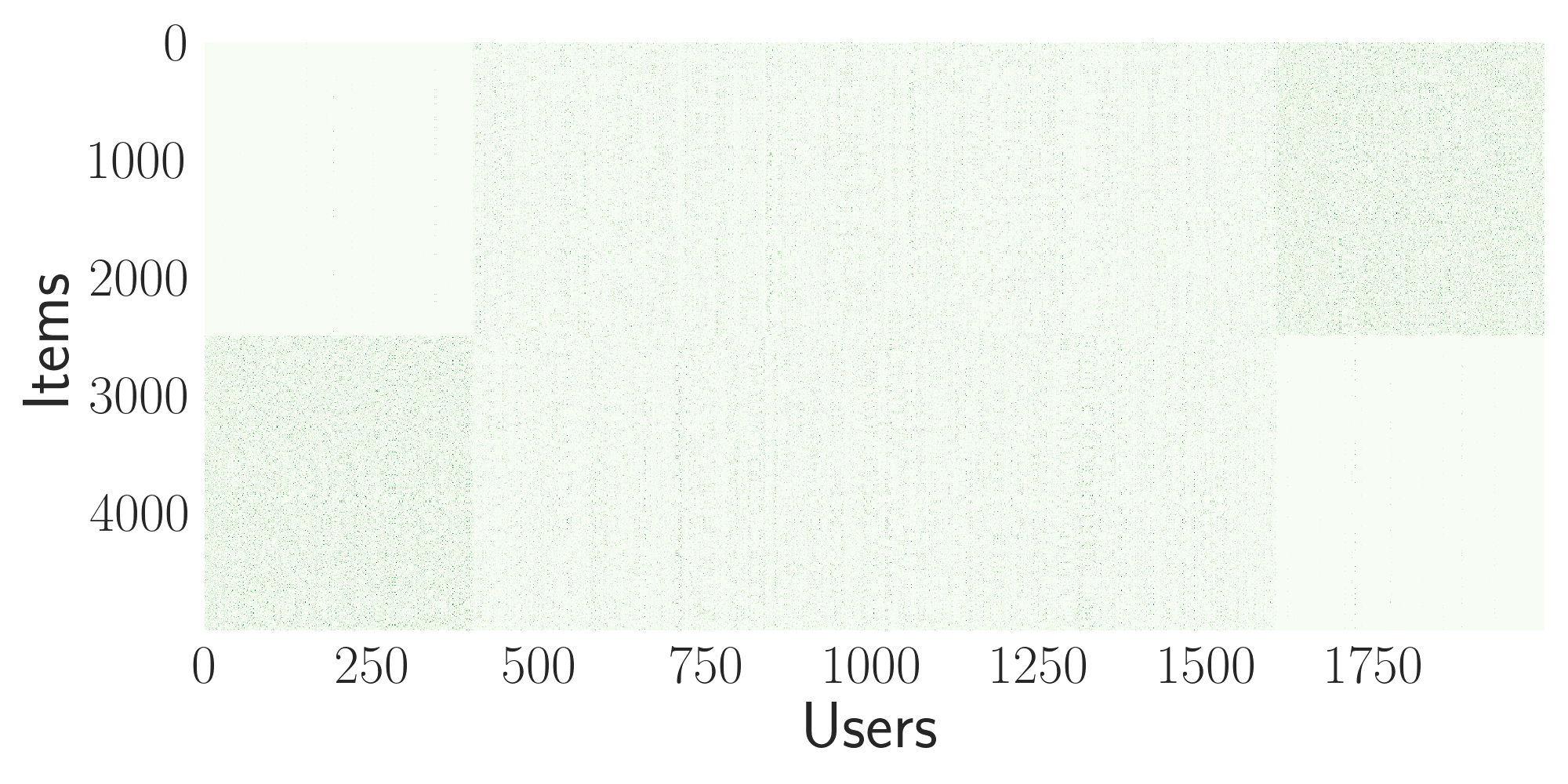}
         \caption{20\%, 60\%, 20\%}
     \end{subfigure}
    \hfill
     \begin{subfigure}[b]{0.32\textwidth}
         \centering
         \includegraphics[width=\textwidth]{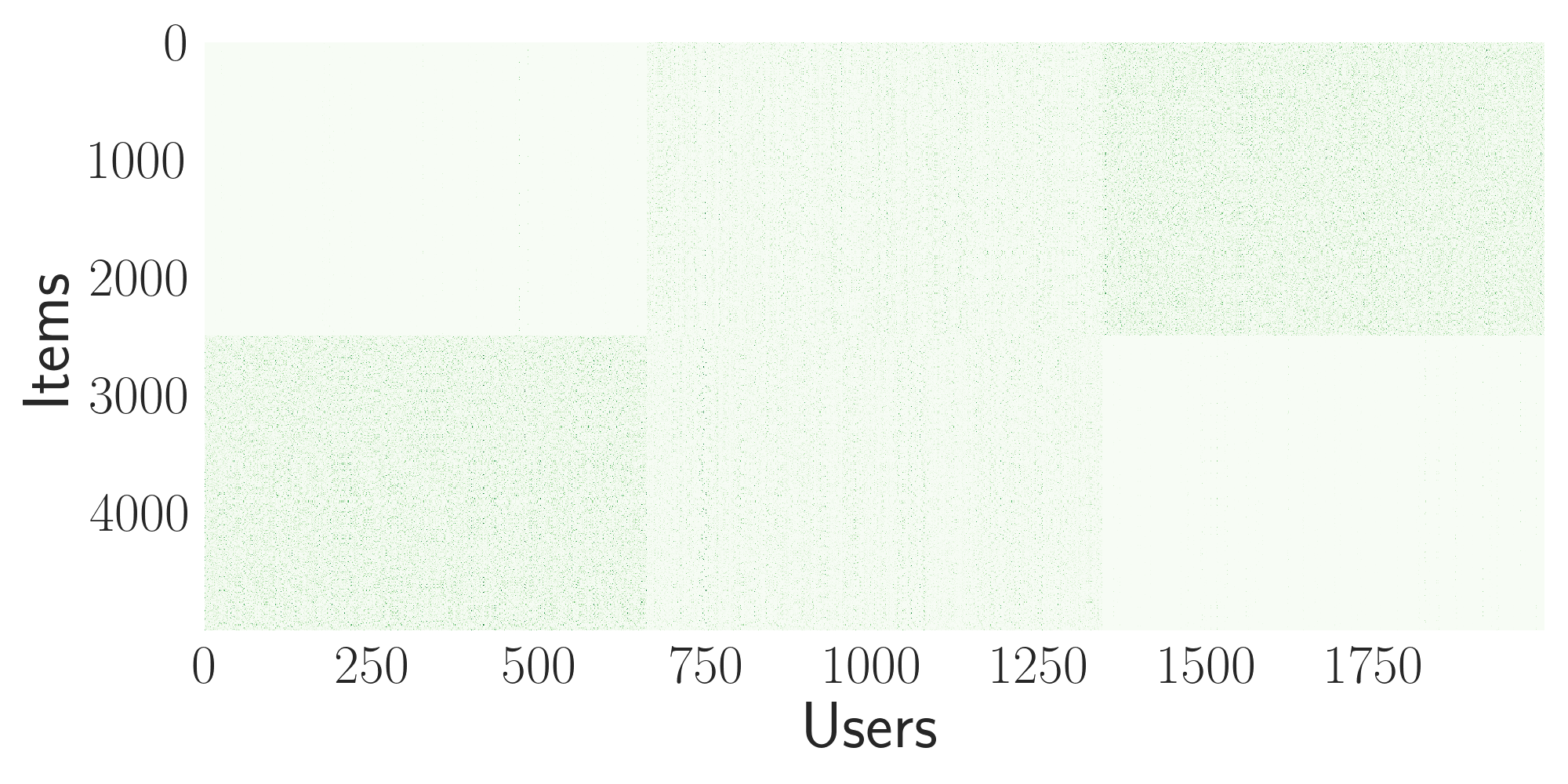}
         \caption{33\%, 33\%, 33\%}
     \end{subfigure}
    \caption{Users' preferences matrix in different samples, varying the proportion of Non-/Semi-/Radicalized users. Assuming that the first (resp. last) $\frac{|I|}{2}$ items are labelled as ``neutral'' (resp. ``harmful''), we impose the Non-Radicalized (resp. Radicalized) users prefer neutral (resp. harmful) items. Conversely, we assume semi-radicalized users span from all items' categories, exhibiting common preferences with the two polarized communities.}
    \label{fig:datasets_utilities}
\end{figure}

\begin{figure}[th!]
    \centering
     \begin{subfigure}[b]{0.5\textwidth}
         \centering
         \includegraphics[width=\textwidth]{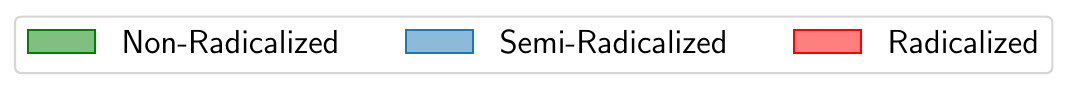}
     \end{subfigure}\\
     \centering
     \begin{subfigure}[b]{0.32\textwidth}
         \centering
         \includegraphics[width=\textwidth]{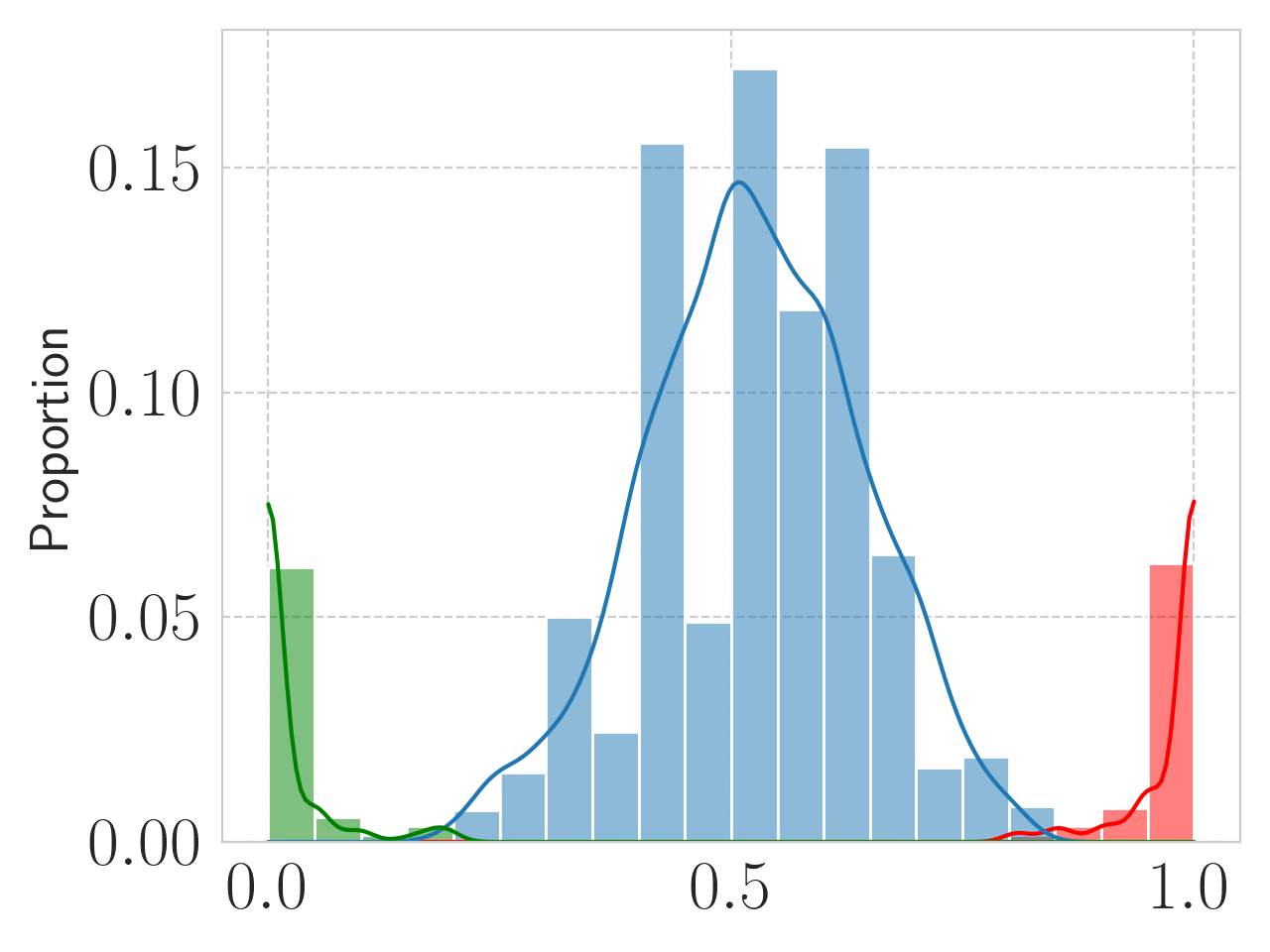}
         \caption{5\%, 90\%, 5\%}
     \end{subfigure}
    \hfill
     \begin{subfigure}[b]{0.32\textwidth}
         \centering
         \includegraphics[width=\textwidth]{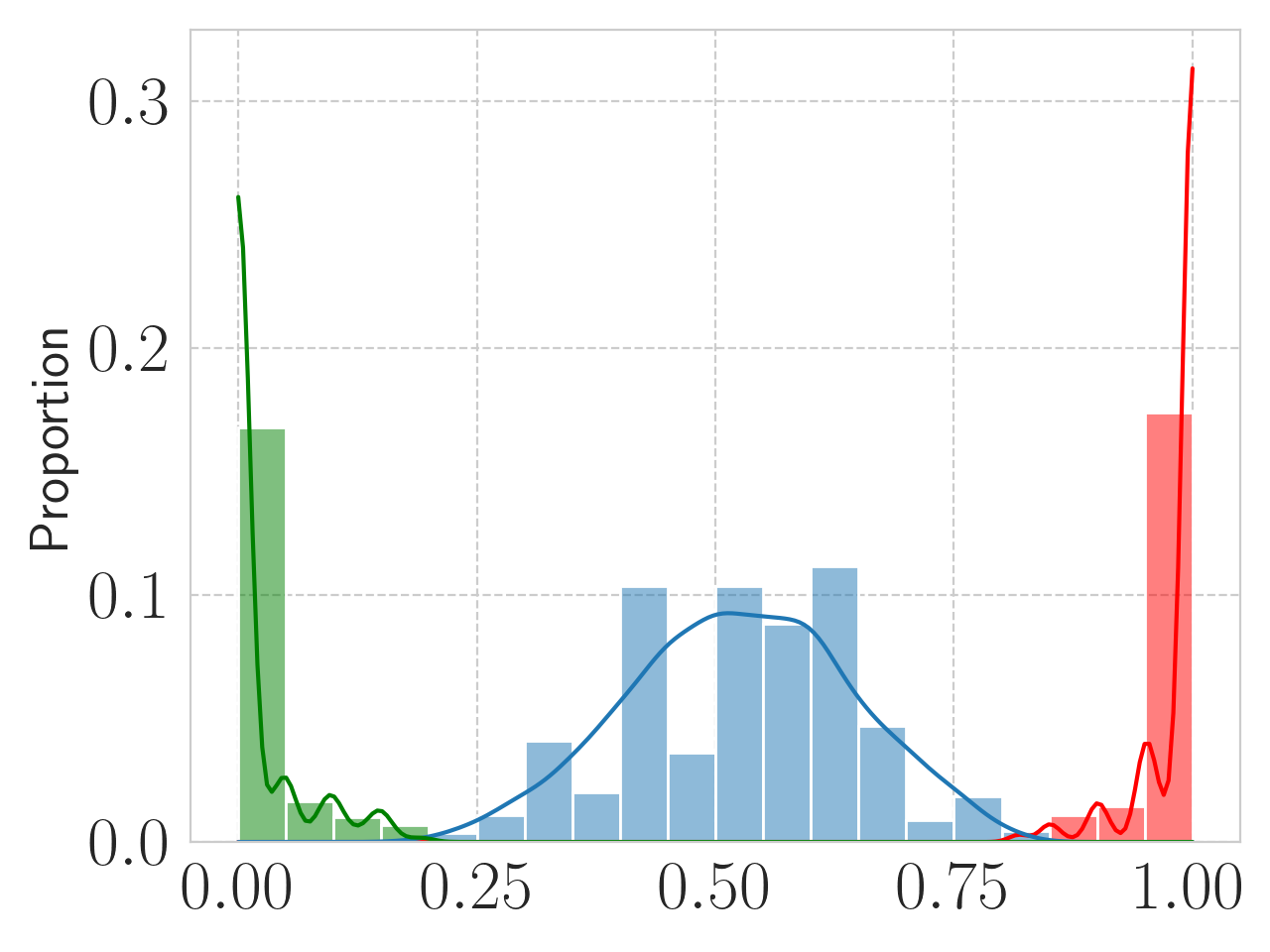}
         \caption{20\%, 60\%, 20\%}
     \end{subfigure}
    \hfill
     \begin{subfigure}[b]{0.32\textwidth}
         \centering
         \includegraphics[width=\textwidth]{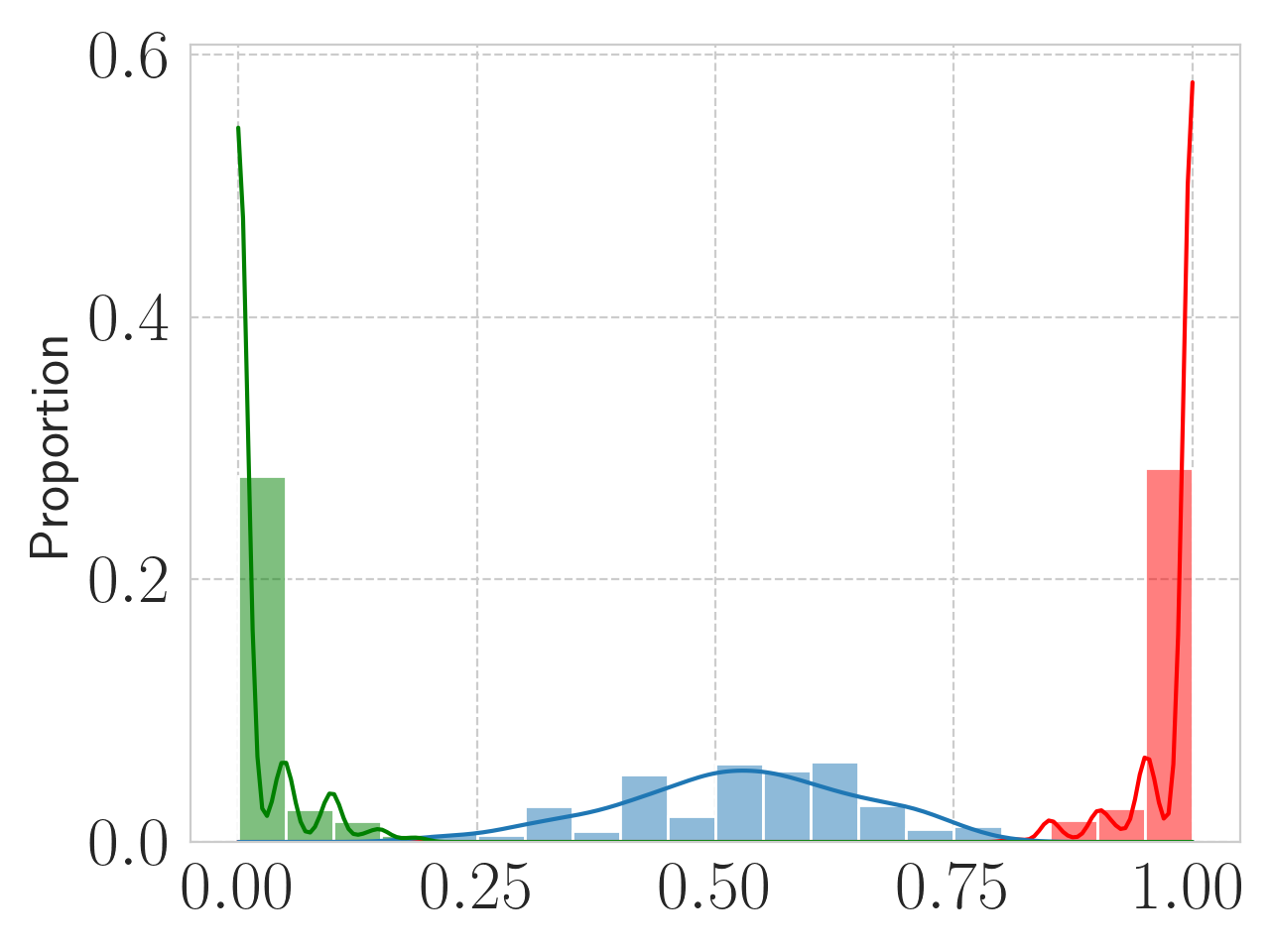}
         \caption{33\%, 33\%, 33\%}
     \end{subfigure}
    \caption{Harmful distribution of Non-/Semi-/Radicalized users, varying the population proportion in the sample. The X-axis represents the harmful percentage in users history, while the Y-axis shows the percentage of users in the dataset.}
    \label{fig:data harmful distribution}
\end{figure}

To validate the effectiveness of our framework, we employ synthetically generated data, which allows us to reproduce different scenarios with great flexibility at low cost~\citep{kronecker, sparse-probabilistic-model, chaneyHowAlgorithmicConfounding2018}.
Specifically, we rely on a synthetic procedure introduced by \cite{DBLP:journals/corr/abs-2407-16594}, where users and items latent features are sampled from a Dirichlet distribution, while Long-Tail distributions are employed to mimic items' popularity and users' engagement. Finally, the likelihood of an interaction between a user $u$ and an item $i$ is proportional to the dot product of the two latent vectors, combined with popularity of $i$ and engagement of $u$.

The aforesaid procedure provides great customizability in terms of both data distributions and users/items group partition.
First, we impose users and items to follow power-law patterns, with parameters $\alpha=2.2$ and $\alpha=2.0$, respectively.
Further, as aforesaid, we split items into two categories: harmful and neutral, respectively, and users into non-radicalized, semi-radicalized, and radicalized, according to the following assumptions: a non-radicalized user shows a percentage of harmful interactions in the range $[0, 0.2]$; a semi-radicalized user, a percentage in the range $(0.2, 0.8)$; a radicalized user, a percentage in the range $[0.8, 1]$. 

Finally, we ensure that the polarized communities (i.e., non-radicalized and radicalized users) share a certain amount of interactions with the semi-radicalized sub-population, by using the parameters setting as suggested by~\citet{DBLP:journals/corr/abs-2407-16594}. 

To validate the robustness of our methodology, we generate three data samples exhibiting an increasing portion of \textit{semi-radicalized} users: (\textit{i}) $5\%, 90\%, 5\%$, (\textit{ii}) $20\%, 60\%, 20\%$, and (\textit{iii}) $33\%, 33\%, 33\%$, respectively for non-radicalized, semi-radicalized and radicalized users. In all the samples, items are equally distributed between harmful and neutral. We fixed the number of users to $2000$, and $|\mathcal{{D}}_u| \geq 20$.

The final result in terms of users' preferences and items distribution are depicted in Figures~\ref{fig:datasets_utilities} and~\ref{fig:data harmful distribution}, respectively.
Notably, despite the preference sets of non-radicalized and radicalized are mostly disjoint, they indeed share interests with the semi-radicalized group, which is supposed to act as a sort of ``\textit{bridge}" between the two sub-populations. 
Therefore, the core intuition is that an higher portion of semi-radicalized users in the sample will trigger the collaborative filtering nature of the recommender algorithm, thus increasing the probability of polarized sub-groups of modifying their initial preferences. In other words, we expect that, the higher the portion of semi-radicalized users in the sample, the stronger the ``algorithmic drift" effect observed.  

\subsection{Settings}\label{sec:settings}
For the experiments, we adopt RecVAE, a collaborative filtering algorithm based on Variational Auto Encoder (VAE) and developed by~\citet{shenbin2020recvae}. The model has been implemented using the Recbole library \citep{recbole}. We trained the recommender system for $100$ epochs setting a hidden dimension equal to $512$. To obtain training, validation, and test sets, the dataset has been split with a ratio of 80/10/10. 
All the code and data used to perform the experiments are publicly available\footnote{\url{https://anonymous.4open.science/r/AlgorithmicDrift-D553}}.

\subsection{Results}
\begin{figure}[th!]
    \centering
     \begin{subfigure}[b]{0.5\textwidth}
         \centering
         \includegraphics[width=\textwidth]{fig/legend_ridge_plot.pdf}
     \end{subfigure}\\
     \centering
     \begin{subfigure}[b]{0.325\textwidth}
         \centering
         \includegraphics[width=\textwidth]{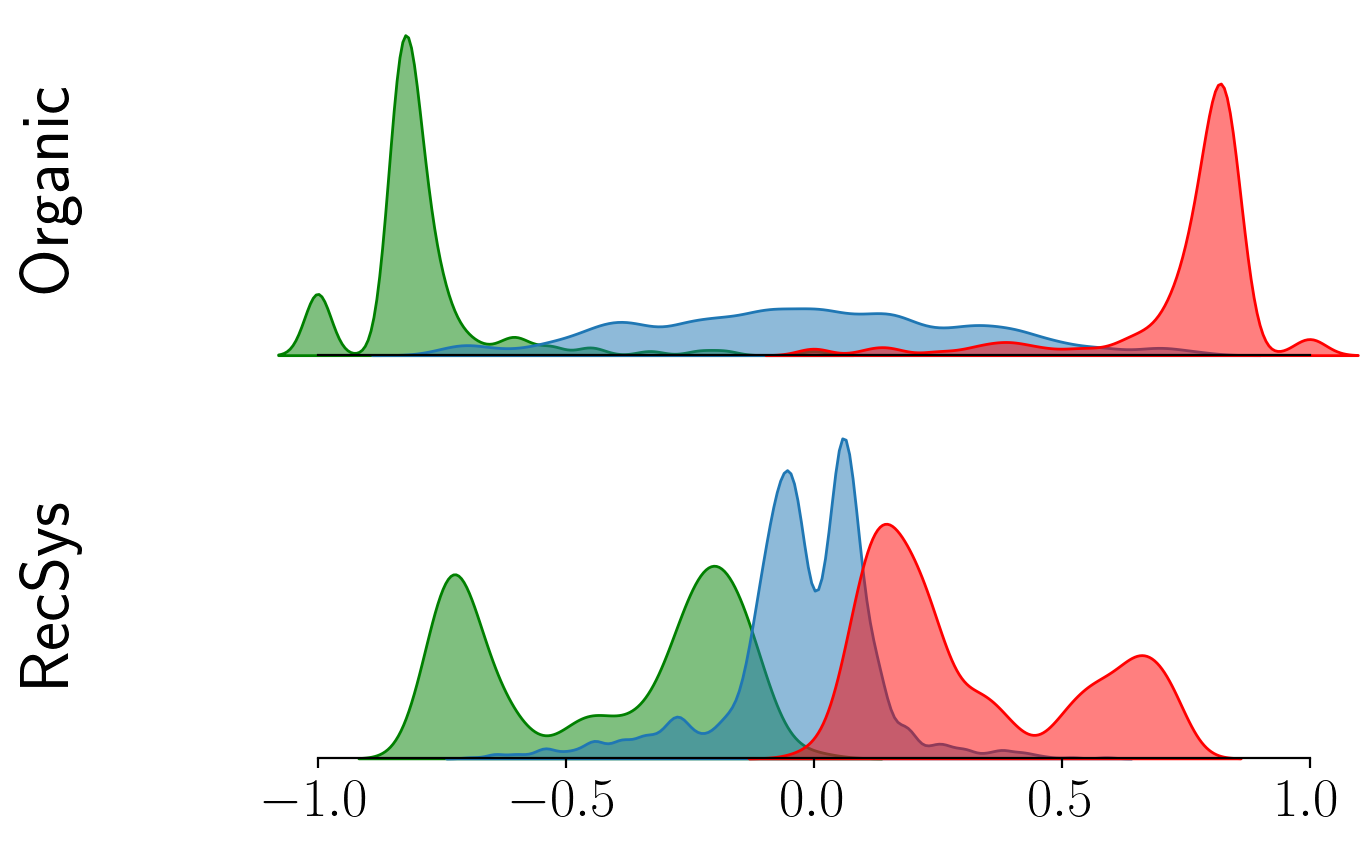}
         \caption{5\%, 90\%, 5\%}
     \end{subfigure}
    \hfill
     \begin{subfigure}[b]{0.325\textwidth}
         \centering
         \includegraphics[width=\textwidth]{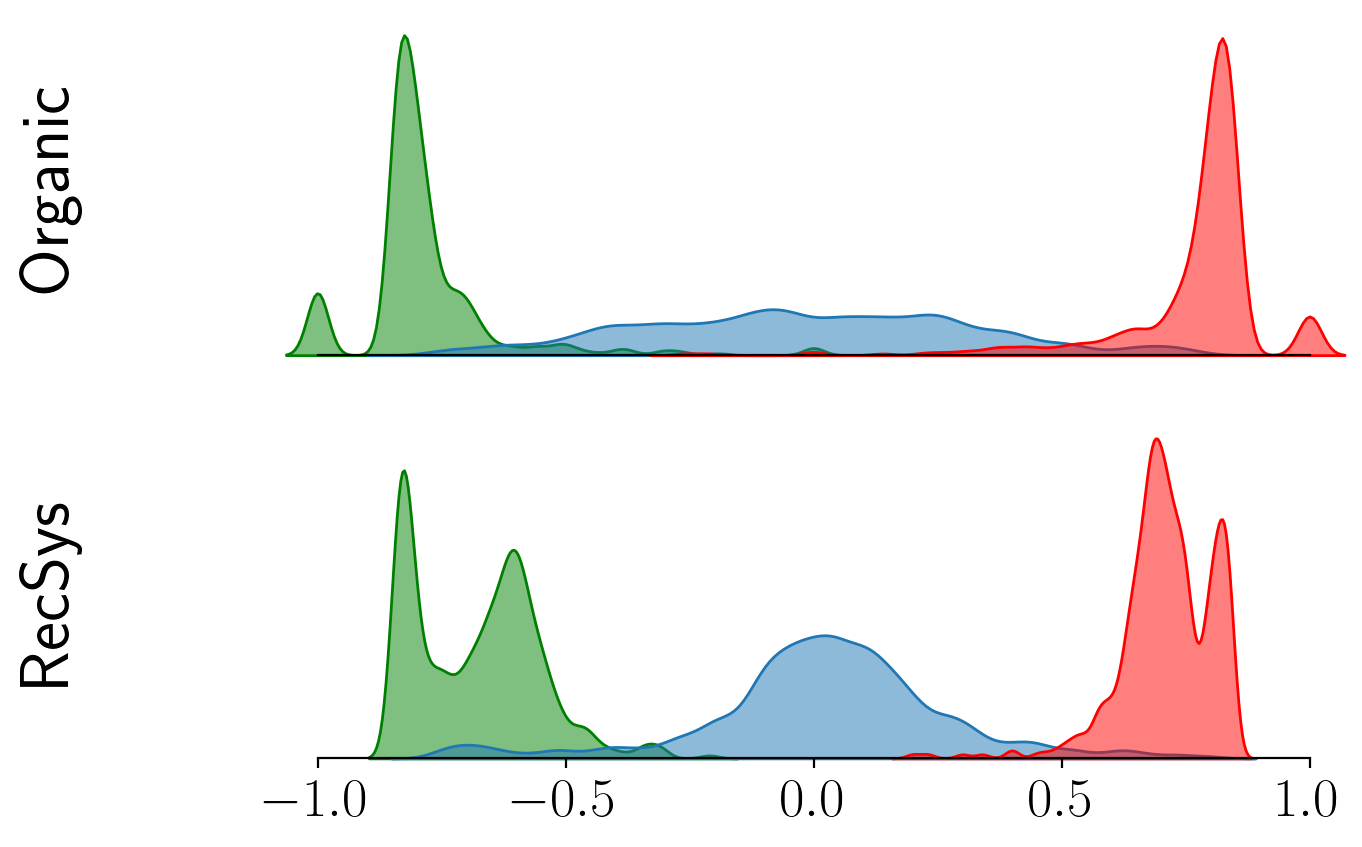}
         \caption{20\%, 60\%, 20\%}
     \end{subfigure}
    \hfill
     \begin{subfigure}[b]{0.325\textwidth}
         \centering
         \includegraphics[width=\textwidth]{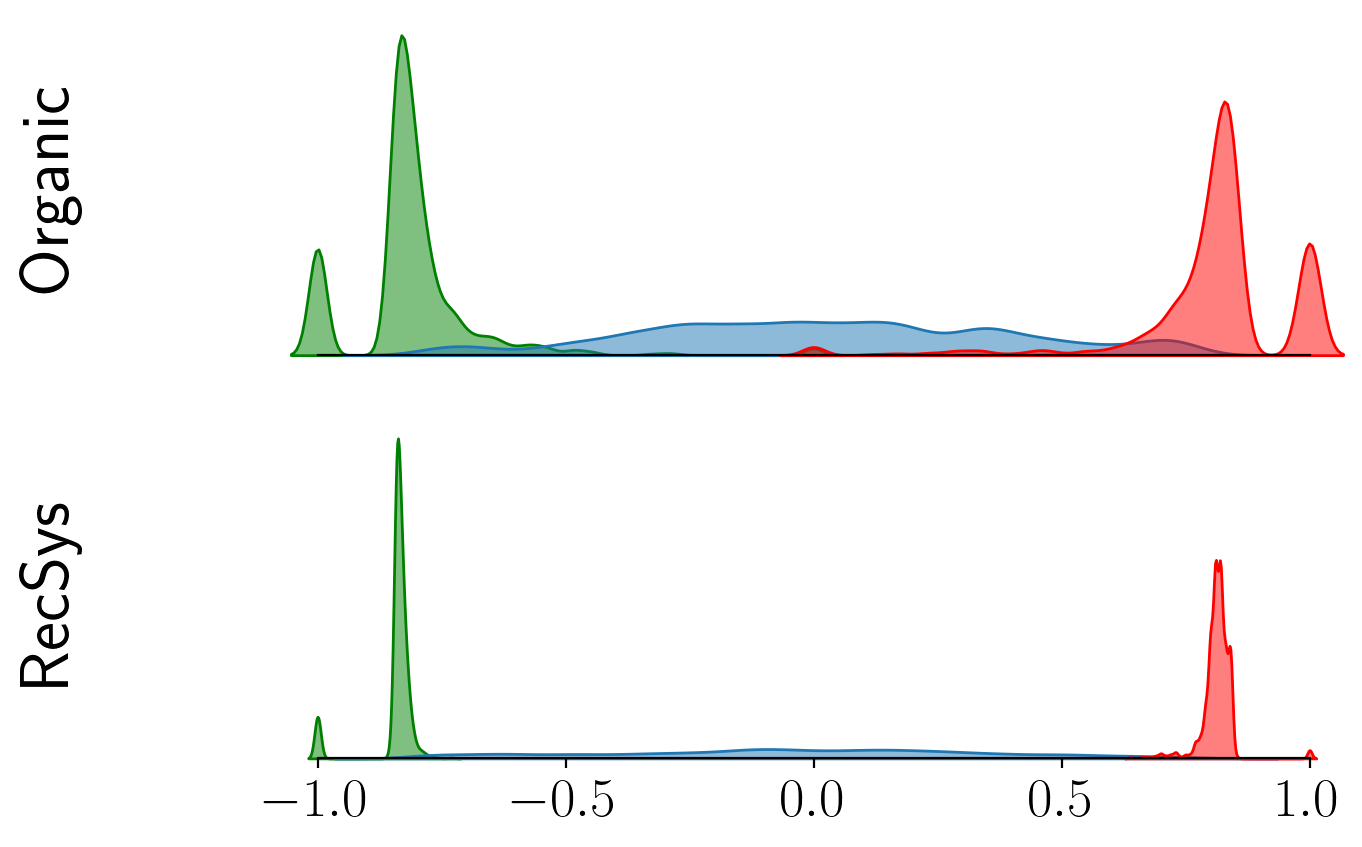}
         \caption{33\%, 33\%, 33\%}
     \end{subfigure}
    \caption{Algorithmic Drift Score (ADS) computed over users graphs by varying the proportion of the starting population (Non-/Semi-/Radicalized \%).}
    \label{fig:ADS-ridge-plot}
\end{figure}

\begin{figure}[th!]
     \centering
     \begin{subfigure}[b]{0.49\textwidth}
         \centering
         \includegraphics[width=\textwidth]{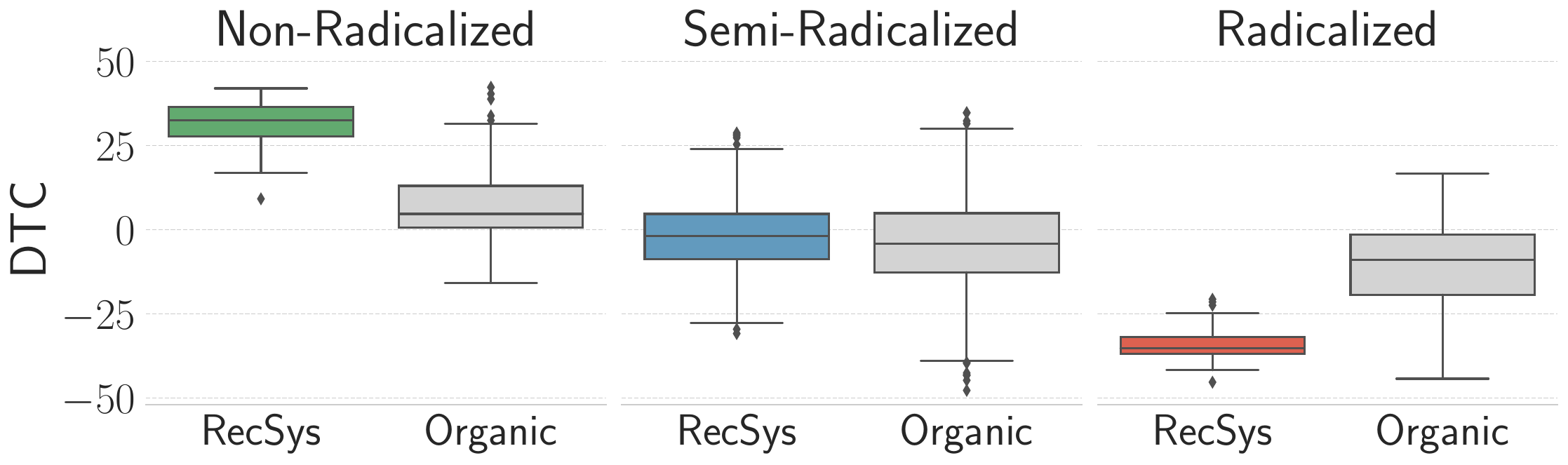}
         \caption{5\%, 90\%, 5\%}
     \end{subfigure}
    \hfill
    \begin{subfigure}[b]{0.49\textwidth}
         \centering
         \includegraphics[width=\textwidth]{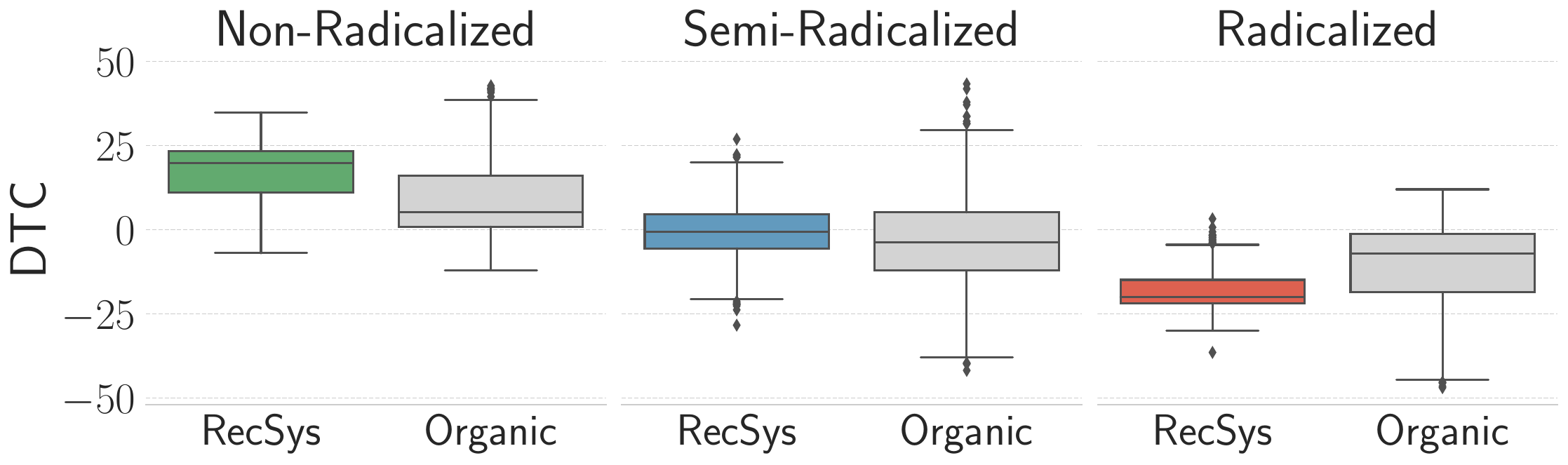}
         \caption{20\%, 60\%, 20\%}
     \end{subfigure}
     \begin{subfigure}[b]{0.49\textwidth}
         \centering
         \includegraphics[width=\textwidth]{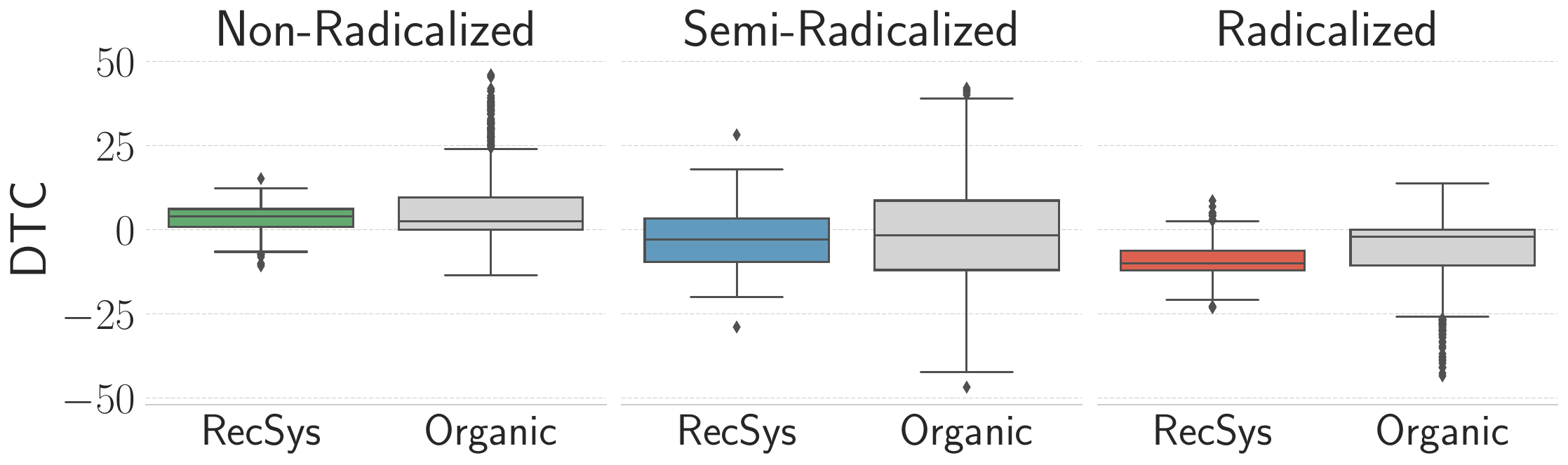}
         \caption{33\%, 33\%, 33\%}
     \end{subfigure}
    \caption{Delta Target Consumption (DTC), expressed in percentage, computed by varying the proportion of the starting population (Non-/Semi-/Radicalized \%).} 
    \label{fig:dhc boxplot}
\end{figure}

\spara{Varying population \textit{proportion}.} First, we investigate the impact of the semi-radicalized sub-population in the sample, which (by construction) is supposed to act as a sort of ``bridge'' between the two polarized communities, by triggering the collaborative filtering nature of the underlying algorithm, and thus fostering the drift effect. To assess if the proposed framework and metrics are able to capture this effect, we fix the resistance ($\gamma$) and inertia ($\delta$) parameter across the three sub-populations, and compare the results with the organic model, i.e., the natural evolution of users preferences without the intervention of any recommender system. In particular, we set $\gamma = 0.1$ and $\delta = 0.5$. Figures~\ref{fig:ADS-ridge-plot} and~\ref{fig:dhc boxplot} show the results in terms of Algorihmic Drift Score (ADS) and Delta Target Consumption (DTC), respectively. 


As we can see from the plots, while the organic model produces values distributions that are almost constant across the different samples, when employing the recommender system, both our metrics are able to capture the drift induced over the users preferences. Indeed, as expected, when the portion of semi-radicalized users increases, this effect is significantly more prominent, while it is basically absent when the three populations exhibit the same percentage, aligning with the organic distributions. 

Notably, the two metrics provides complementary information with respect to the target category: while the DTC rate quantifies the consumption increment in the final user history, the ADS provides a probabilistic perspective computed over the interactions graph.  

\spara{Impact of \textit{resistance} and \textit{inertia}.} Further, we conduct a more fine-grained analysis on the effects of varying the user behavioral factors $\gamma$ (resistance) and $\delta$ (inertia).

As mentioned in Sections~\ref{sec:intro} and~\ref{sec:user_behaviour}, we denote the resistance as the user hesitancy in relying on the recommender: the higher, the more prone the user is to autonomously select an item from the catalog; conversely, the inertia models the user tendency in following the recommendations: the higher, the more trust the user shows in the algorithm when choosing the item, ignoring their own preferences. 
\begin{figure}[th!]
     \centering
     \begin{subfigure}[b]{0.32\textwidth}
         \centering
         \includegraphics[width=\textwidth]{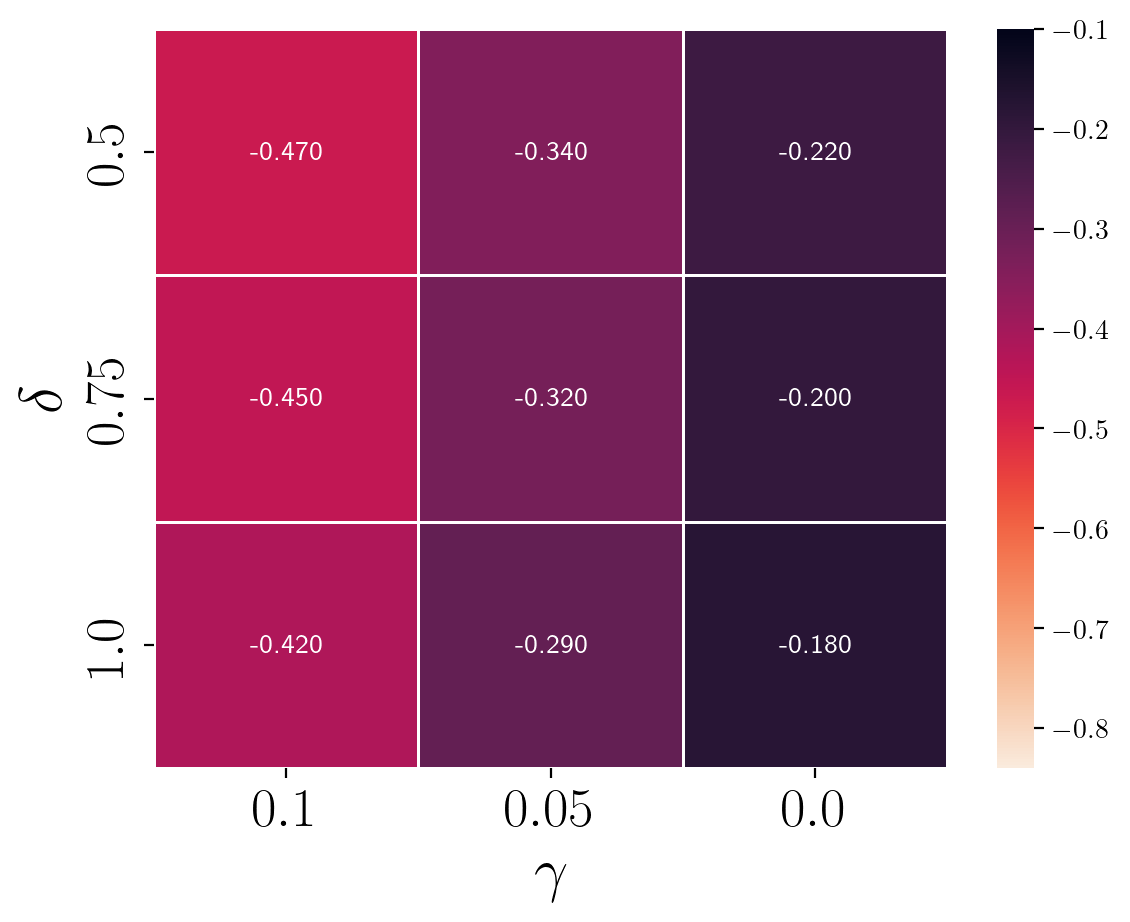}
     \end{subfigure}
    \hfill
    \begin{subfigure}[b]{0.32\textwidth}
         \centering
         \includegraphics[width=\textwidth]{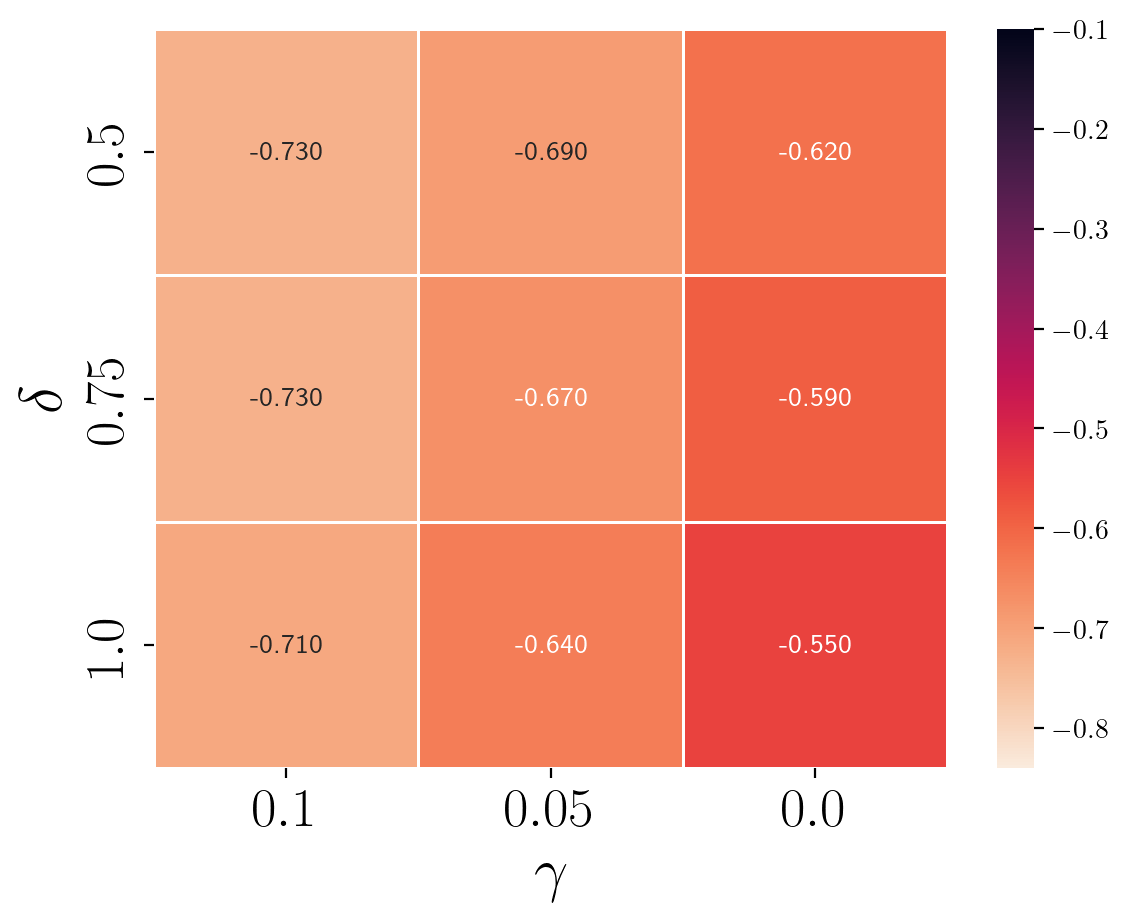}
     \end{subfigure}
    \hfill
     \begin{subfigure}[b]{0.32\textwidth}
         \centering
         \includegraphics[width=\textwidth]{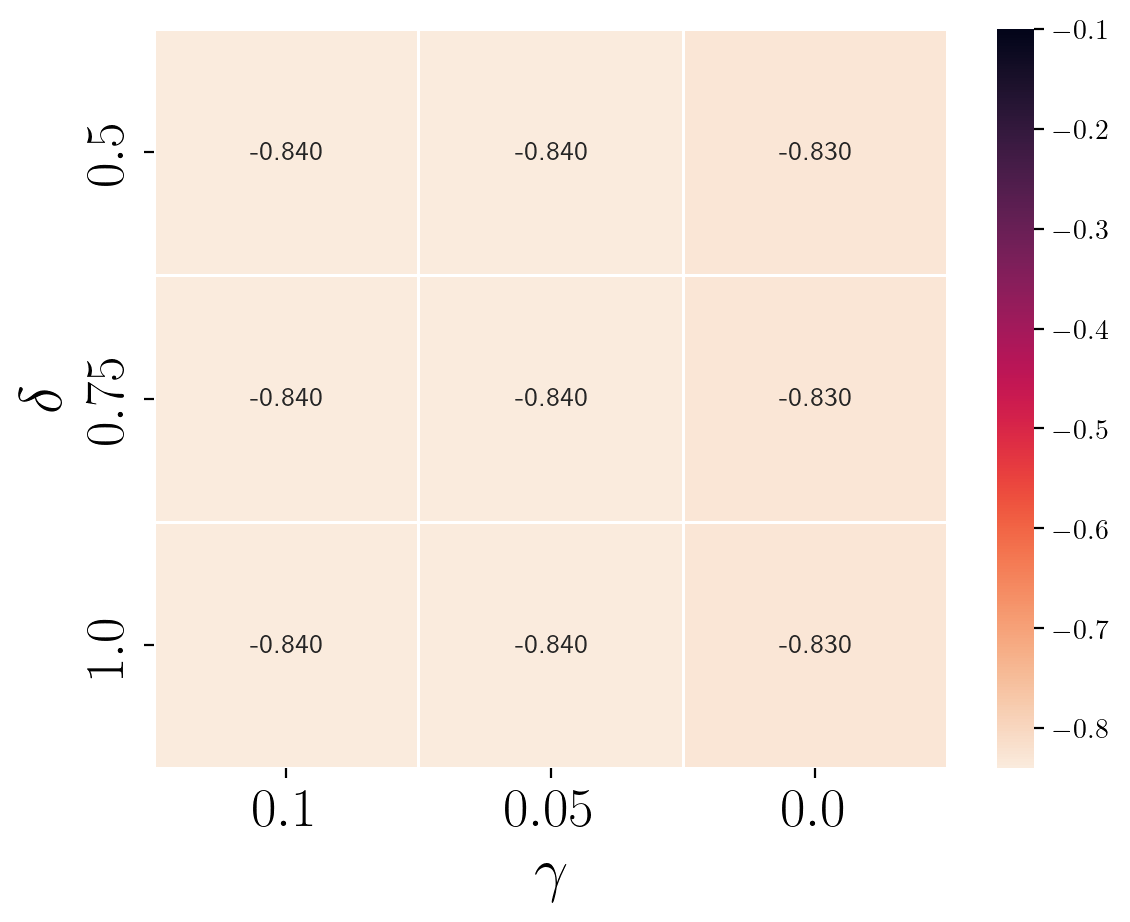}
     \end{subfigure}
     \begin{subfigure}[b]{0.32\textwidth}
         \centering
         \includegraphics[width=\textwidth]{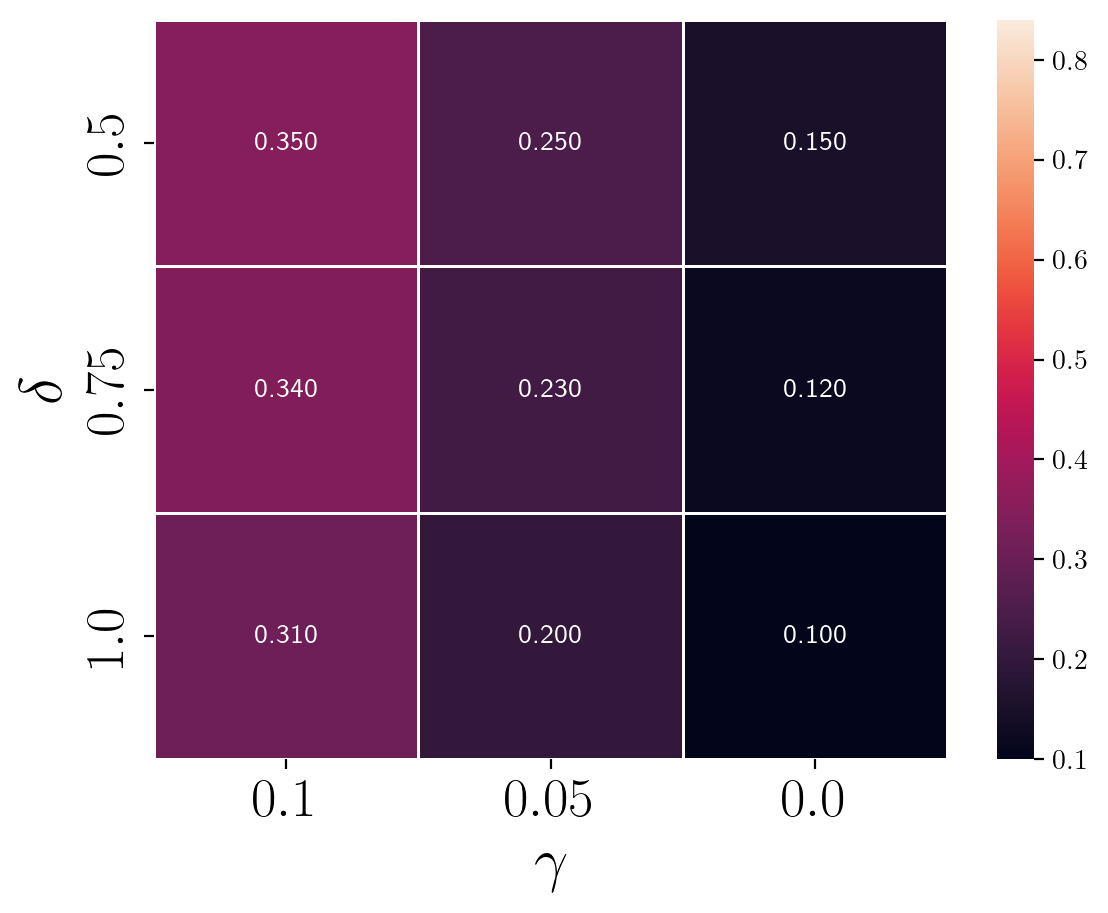}
         \caption{Population proportion 5\%, 90\%, 5\%}
     \end{subfigure}
    \hfill
    \begin{subfigure}[b]{0.32\textwidth}
         \centering
         \includegraphics[width=\textwidth]{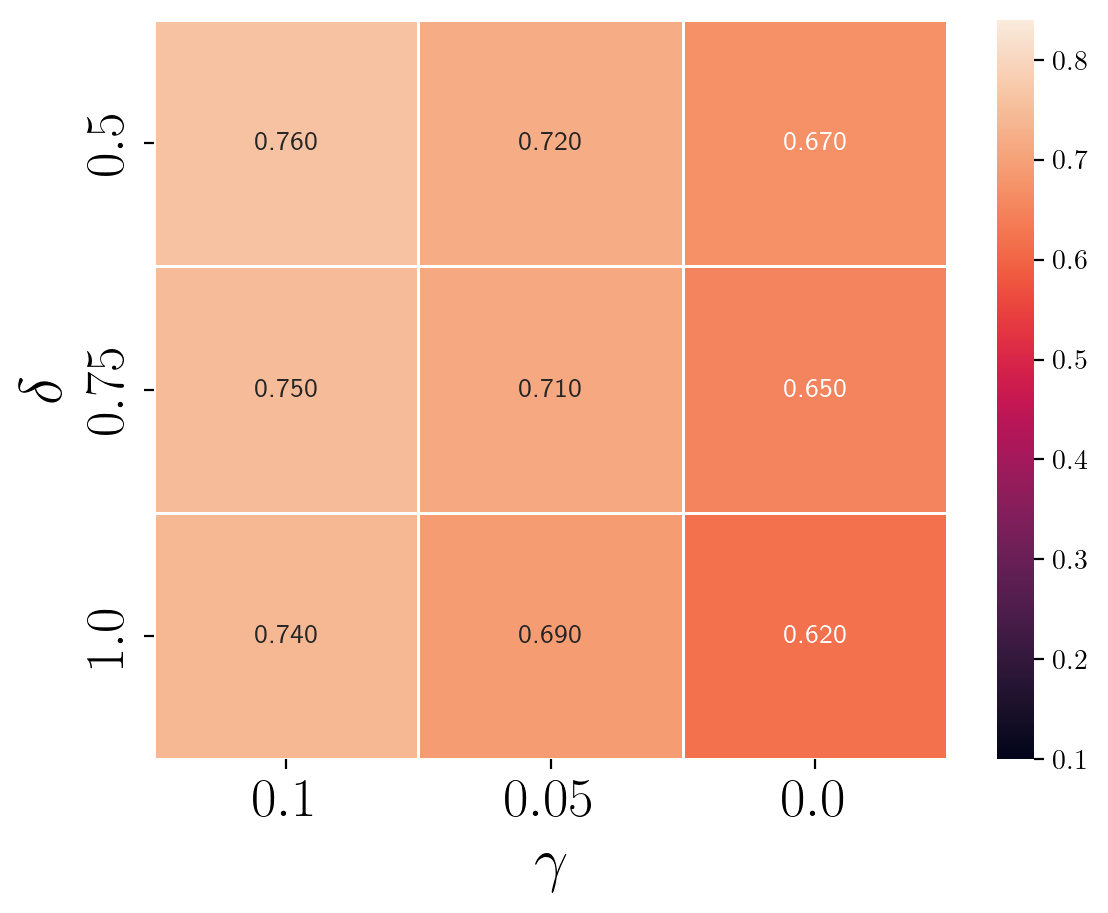}
         \caption{Population proportion 20\%, 60\%, 20\%}
     \end{subfigure}
     \hfill
     \begin{subfigure}[b]{0.32\textwidth}
         \centering
         \includegraphics[width=\textwidth]{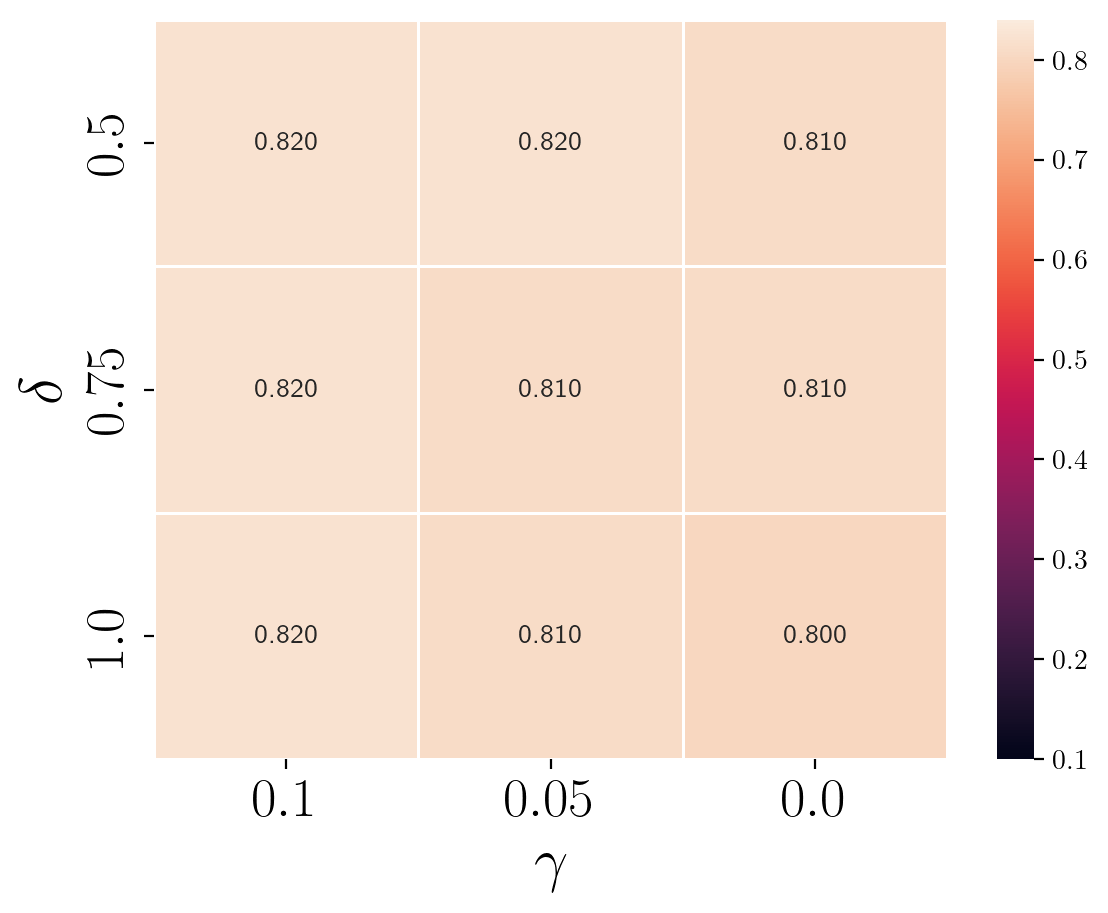}
         \caption{Population proportion 33\%, 33\%, 33\%}
     \end{subfigure}
    \caption{Algorithmic Drift Score (ADS) induced by the recommendation algorithm over non-radicalized (top row) and radicalized (bottom row) by varying the proportion of the starting population.}
    \label{fig:ads heatmap}
\end{figure}
\begin{figure}[th!]
     \centering
     \begin{subfigure}[b]{0.325\textwidth}
         \centering
         \includegraphics[width=\textwidth]{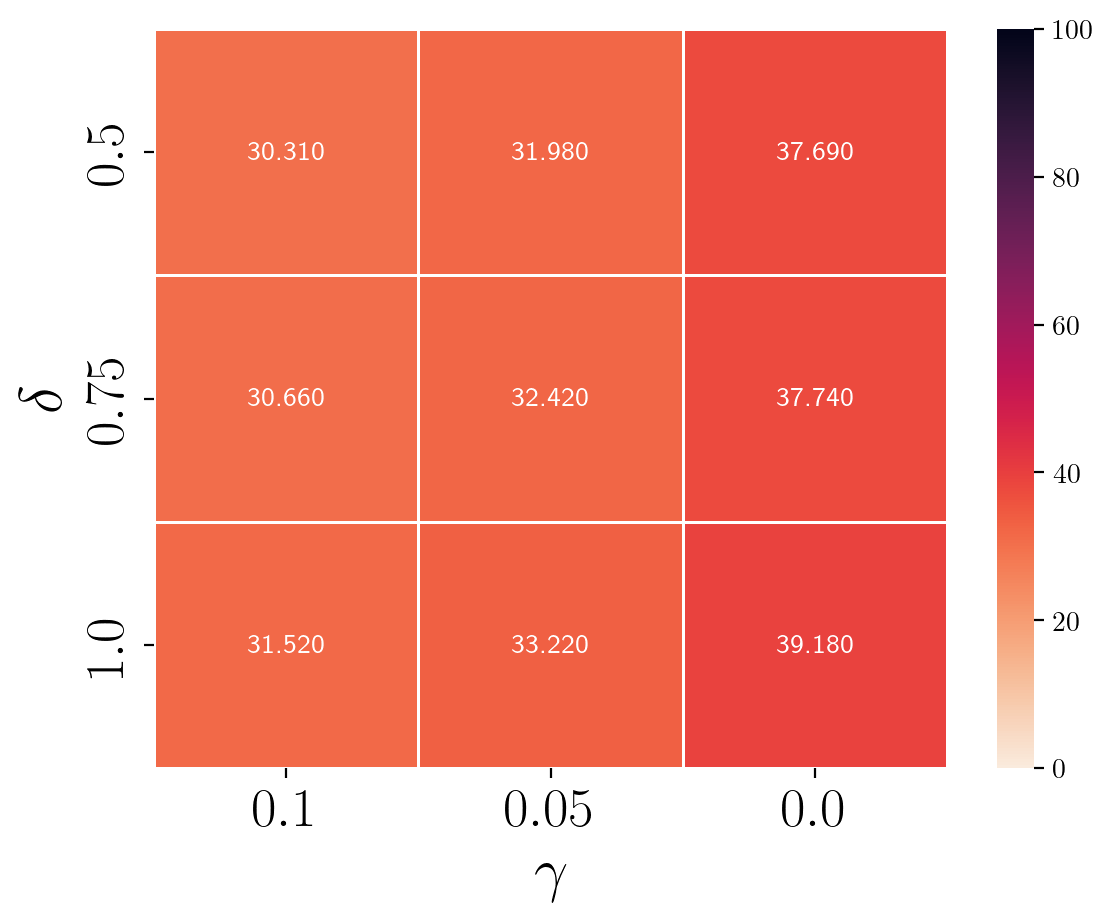}
     \end{subfigure}
    \begin{subfigure}[b]{0.325\textwidth}
         \centering
         \includegraphics[width=\textwidth]{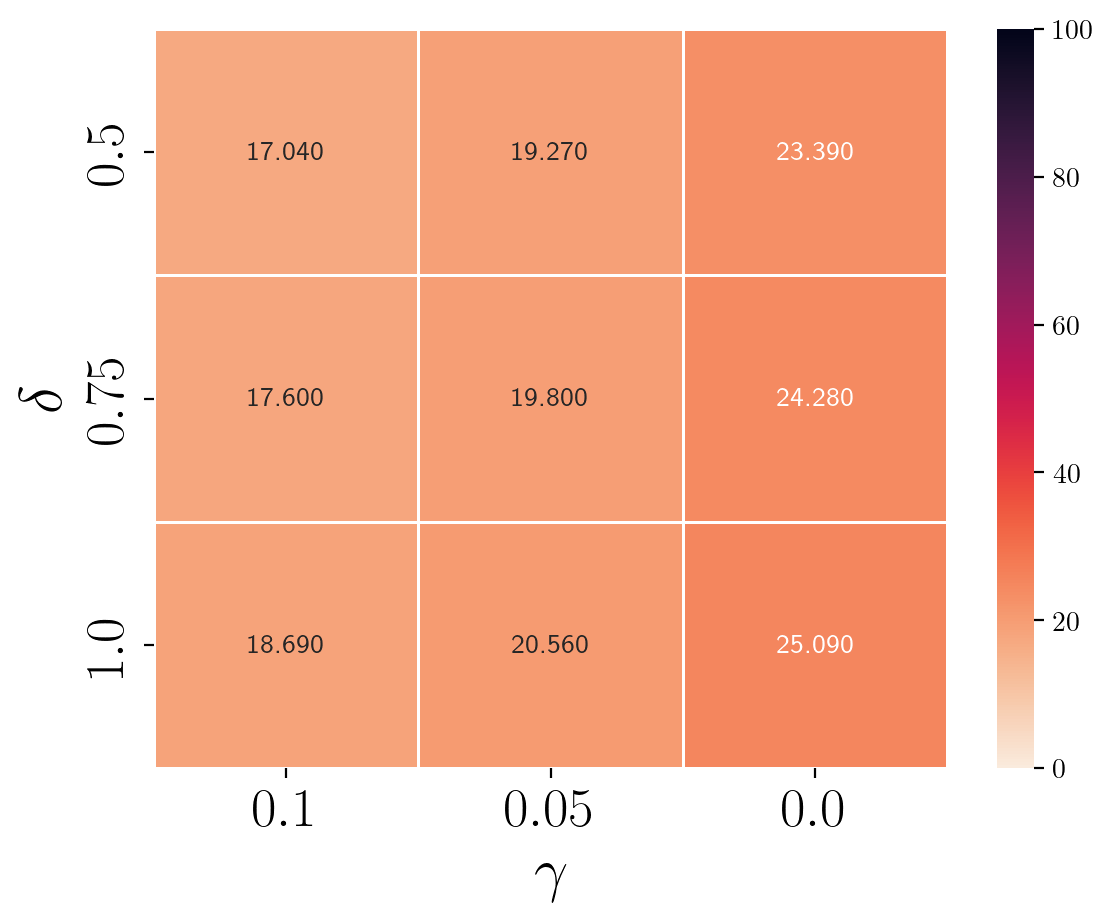}
     \end{subfigure}
     \begin{subfigure}[b]{0.325\textwidth}
         \centering
         \includegraphics[width=\textwidth]{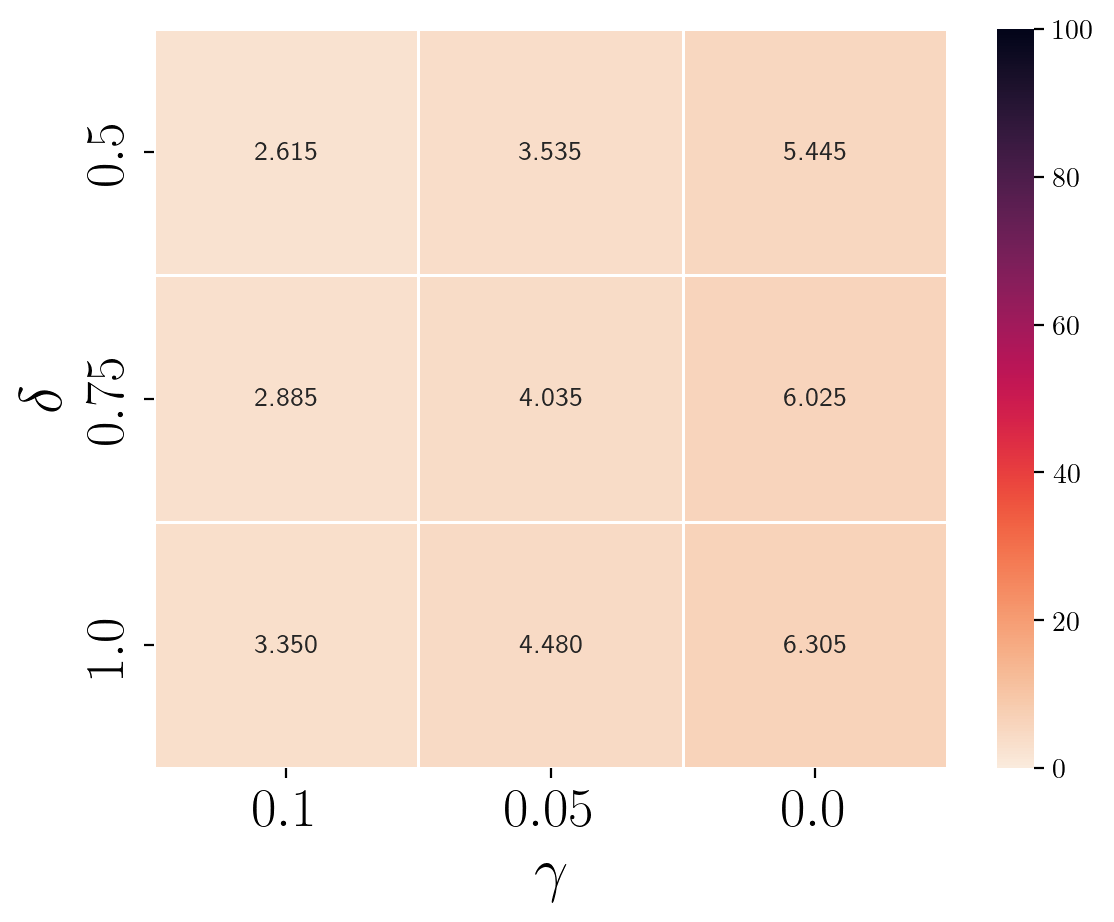}
     \end{subfigure}
    \centering
     \begin{subfigure}[b]{0.325\textwidth}
         \centering
         \includegraphics[width=\textwidth]{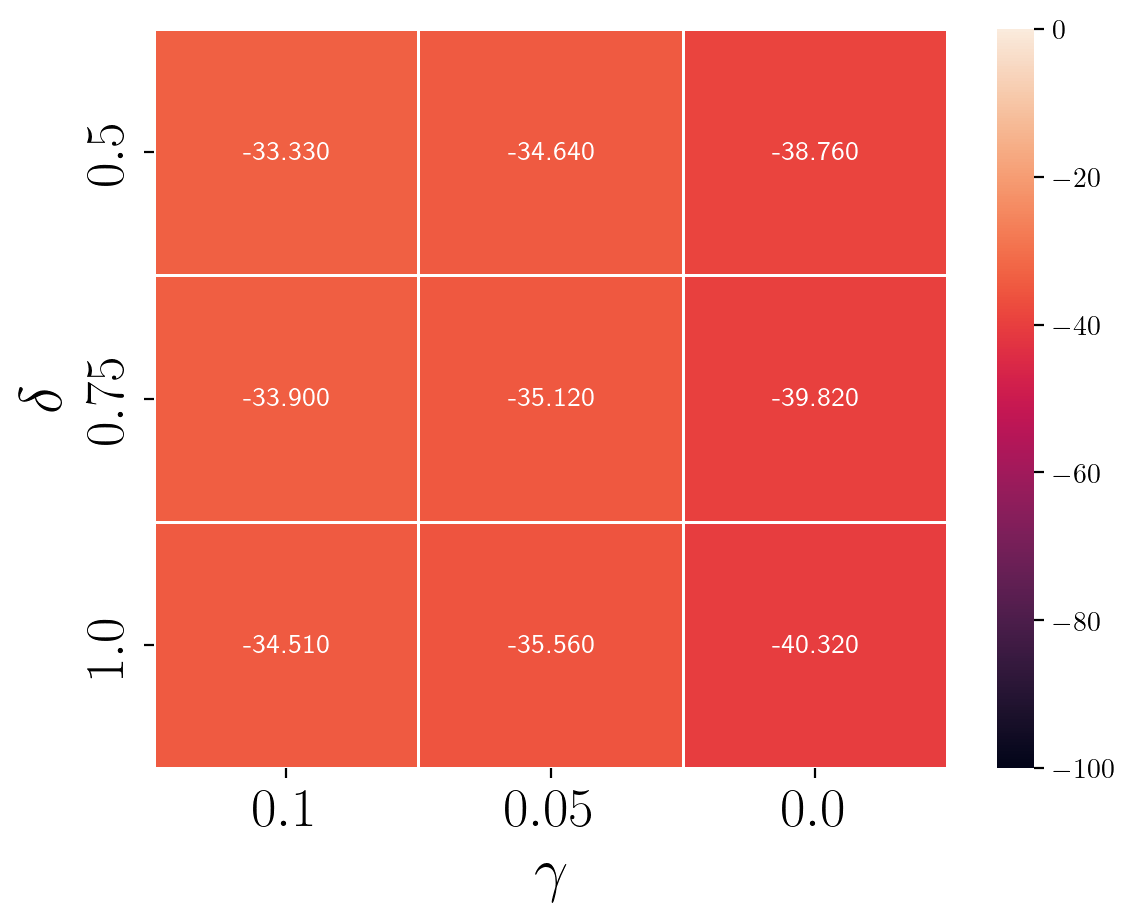}
         \caption{Population proportion 5\%, 90\%, 5\%}
     \end{subfigure}
    \begin{subfigure}[b]{0.325\textwidth}
         \centering
         \includegraphics[width=\textwidth]{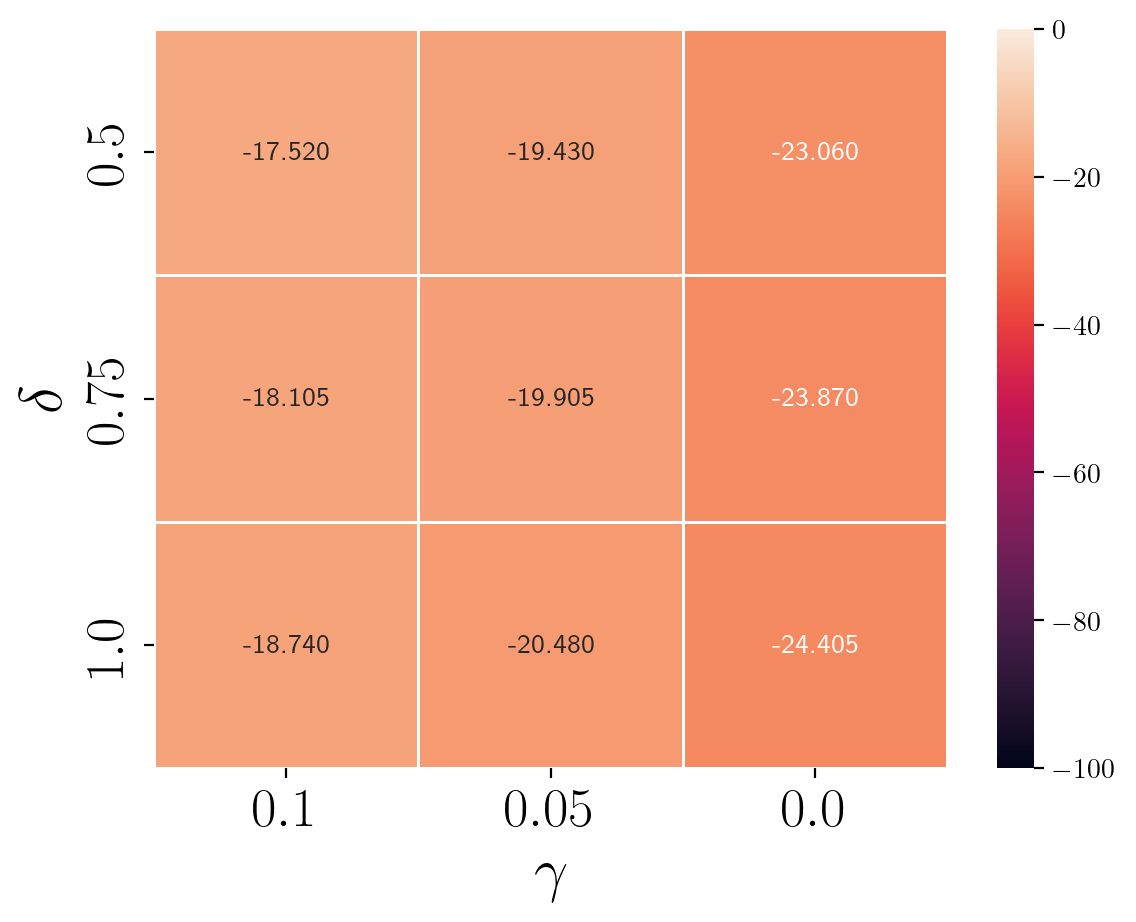}
         \caption{Population proportion 20\%, 60\%, 20\%}
     \end{subfigure}
     \begin{subfigure}[b]{0.325\textwidth}
         \centering
         \includegraphics[width=\textwidth]{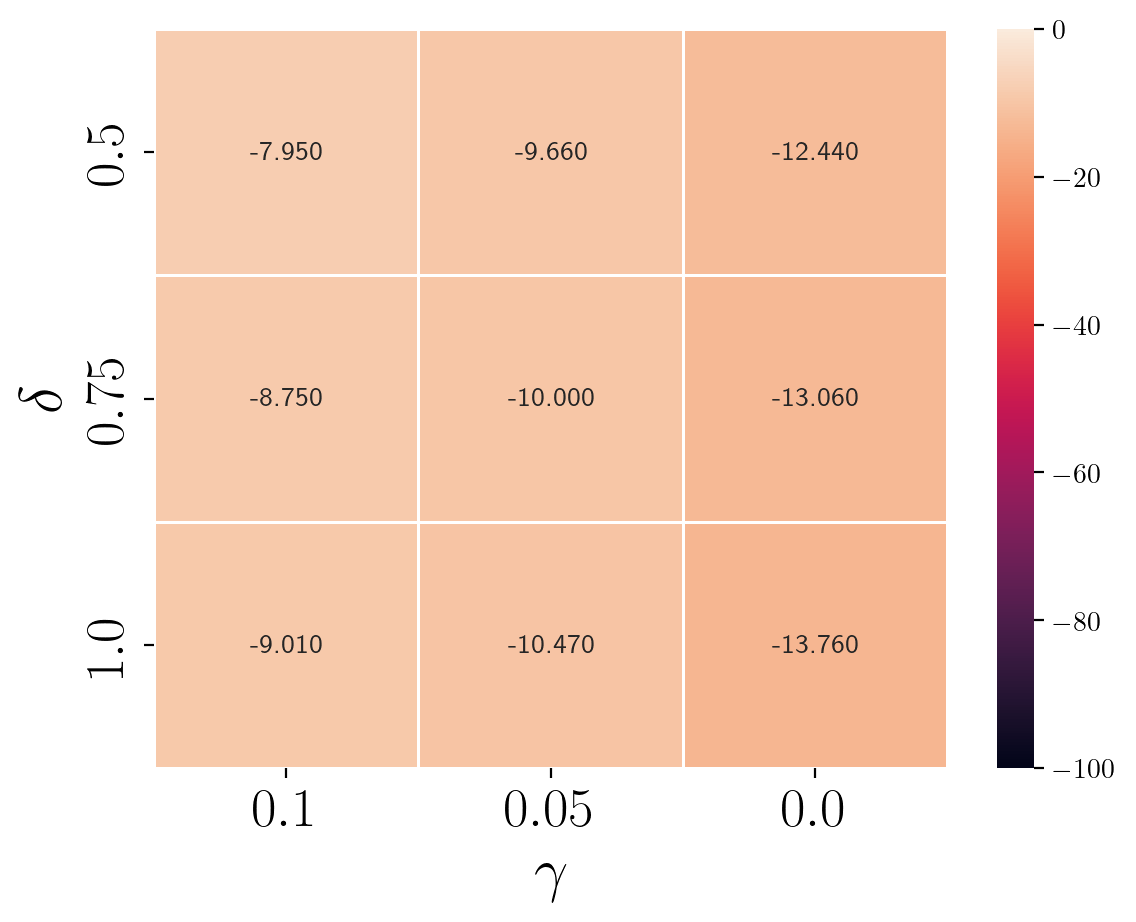}
         \caption{Population proportion 33\%, 33\%, 33\%}
     \end{subfigure}
    \caption{Delta Target Consumption (DTC), in percentage, computed over non-radicalized (top row) and radicalized (bottom row) users by varying the population proportion, the resistance ($\gamma$) and inertia ($\delta$) parameters.}
    \label{fig:dhc heatmap}
\end{figure}

Following from these assumptions, we indeed expect that when $\gamma$ decreases (low resistance), and $\delta$ increases (high inertia), the drift effect is more prominent; vice-versa, when $\gamma$ is high or $\delta$ is low, the deviations of users preferences in the long run are negligible.

For the experiments, we vary $\delta$ in the range $[0.5, 0.75, 1.0]$, and $\gamma$ in the range $[0.0, 0.05, 0.1]$. The results are depicted in Figures~\ref{fig:ads heatmap} and~\ref{fig:dhc heatmap}, in terms of ADS and DTC, respectively. In each figure, the heatmaps in the top-row show the median value of the corresponding metric computed over non-radicalized users, while the heatmaps in the bottom-row refer to the radicalized population. Further, each column indicates a different proportion of the original sample, in terms of non-, semi-, and radicalized users.

Two considerations can be made on the results. First, as in the previous set of experiments, the portion of semi-radicalized users has great impact on the final results: the higher the population, the darker the grid, i.e., the more prominent is the deviation of the initial users preferences. Secondly, fixed a sample proportion, the devised drift effect is increasingly evident when going from $\delta = 0.5, \gamma = 0.1$ (top-left) to $\delta = 1.0, \gamma = 0.0$ (bottom-right), thus perfectly reflecting our intuition.  

\begin{figure}[th!]
     \centering
     \begin{subfigure}[b]{\textwidth}
         \centering
         \includegraphics[width=0.8\textwidth]{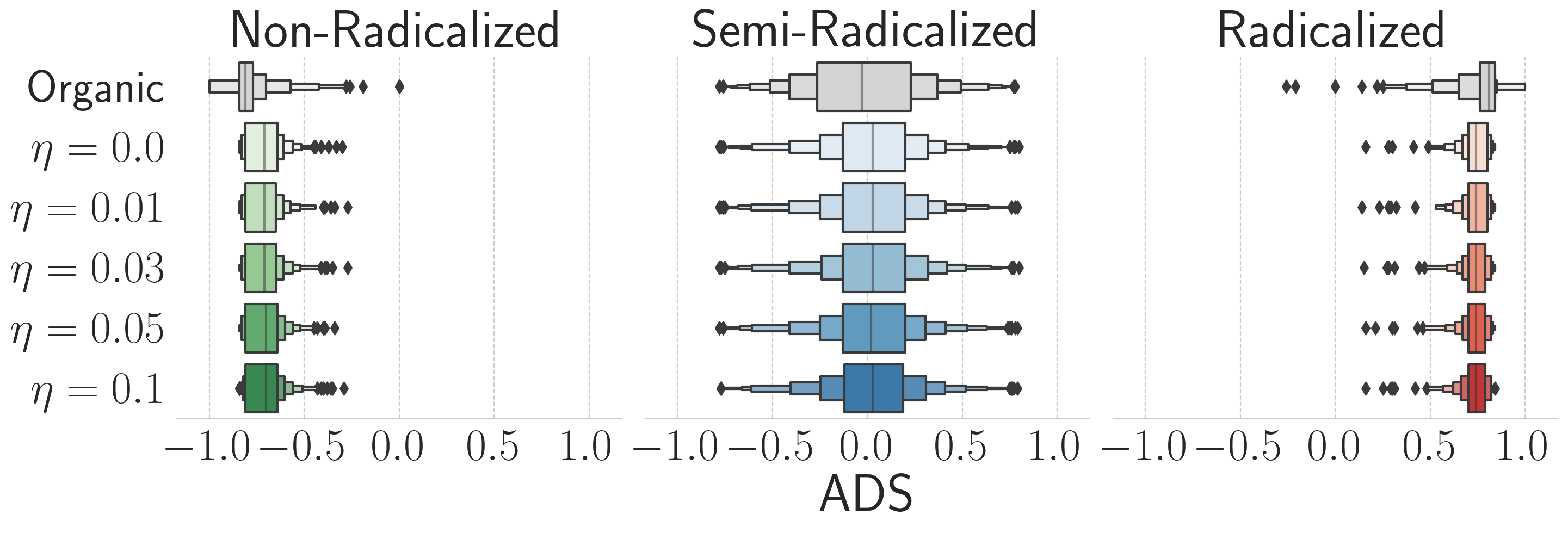}
     \end{subfigure}
     \vfill
     \begin{subfigure}[b]{\textwidth}
         \centering
         \includegraphics[width=0.8\textwidth]{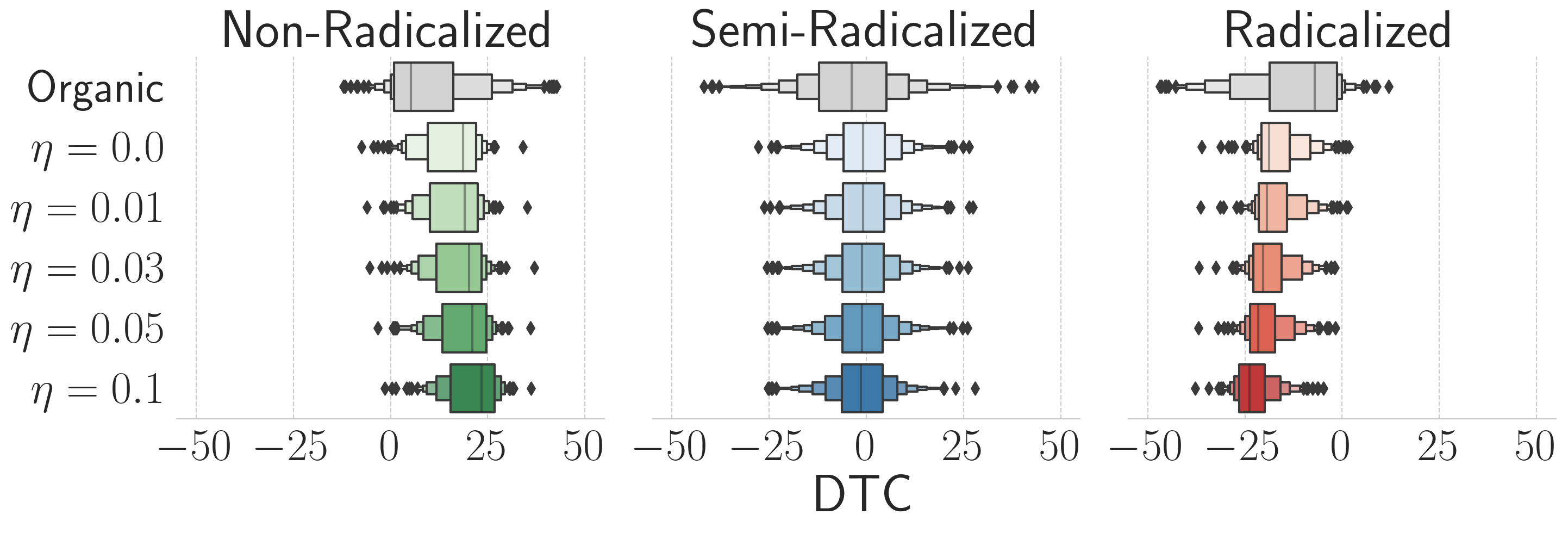}
     \end{subfigure}
     
    \caption{Algorithmic Drift Score (ADS) (top-row) and Delta Target Consumption (DTC, in percentage) (bottom-row) computed varying the random factor $\eta$, comparing with the organic model.} 
    \label{fig:eta random boxplot}
\end{figure}

\spara{Increasing choice \textit{randomness}.} Finally, we aim at evaluating the impact of the random parameter $\eta$ in the user selections, resembling exogenous factors like a friend's suggestion or a misclick. 
Since we assume random factors being very low in practice, for this set of experiments we set $\eta$ spanning in the range [0.01, 0.03, 0.05, 0.1]. We further fix $\gamma = 0.1$ and $\delta = 1.0$, and the population proportion to be equal $20\%, 60\%, 20\%$. 
Intuitively, the spurious interactions introduced by means of randomness should not alter the user preferences in the long term, who will indeed follow their own intrinsic preferences. Figure~\ref{fig:eta random boxplot} shows the experimental results in terms of ADS (top-row) and DTC (bottom-row), by varying the $\eta$ parameter, and comparing to the same setting with no randomness ($\eta = 0$), as well as to the organic model.

Two considerations can be here made, regarding the different effects captured by the two metrics. Indeed, while DTC devises a slight increment of harmful (resp., neutral) consumption by non-radicalized (resp., radicalized) users, this is due to the fact that, the higher $\eta$, the higher the probability the user picks an item belonging to the opposite category w.r.t. their initial history. In other words, if the user belongs to the non-radicalized community, the majority of their initial interactions consists of neutral items, thus increasing the probability of (randomly) selecting content tagged as harmful (and vice-versa, given a radicalized user). 
Remind in fact that, in each of the $B$ independent rounds, we assume an user cannot interact with the same item twice. 

This, however, does not necessarily imply a drift in their natural preferences, i.e., an higher probability in encountering and remaining in harmful (resp., neutral) pathways: indeed, the users distributions computed in terms of ADS are not affected by increasing the $\eta$ parameter, thus showing no alteration in terms of users preferences in the long term, as we expected.

\section{Conclusions and Future Work}
\label{sec:conclusions}
In this paper, we proposed a novel stochastic model for studying potential deviations of users' preferences due to the influence of recommendation systems in the long term. We denote this phenomenon as ``\textit{algorithmic drift}'', and we introduce two novel metrics, namely Algorithmic Drift Score and Delta Target Consumption, in order to quantify it. Further, our framework provides great flexibility in representing user behaviors throughout the simulation process, by modeling behavioral patterns such as user \textit{resistance} to recommendations, \textit{inertia} in following the provided suggestions, and choice \textit{randomness} due to exogenous hence uncontrollable factors.

Our main contributions can be indeed summarized as follows: (\textit{i}) the definition of the \textit{algorithmic drift} concept and the introduction of two novel metrics in order to quantify it; (\textit{ii}) the implementation of a stochastic model for analyzing the impact of recommender systems in the long term; and (\textit{iii}) an extensive evaluation through a practical use-case based on a collaborative-filtering algorithm, showing the model's capabilities across different scenarios. The ultimate result is a robust controlled environment for evaluating the recommendation algorithm before deployment. 

The proposed framework is amenable for further extensions in many different directions. 
First, we assume the items catalog to be fixed, while a more dynamic setting could be considered where novel items are continuously introduced. Also, the proposed user model does not take into account contextualization. In the depicted use case, items are categorized as either harmful or neutral. However, more realistic scenarios can embrace situations where items are tagged as harmful depending on the user features or the recommendation context. Moreover, radicalization and harmfulness can also be considered according to specific ideological axes upon which users and items can be aligned. 
A final line of further investigation is the adaptation of the proposed methodology to study other typical weaknesses that can occur in a recommendation setting, such as popularity bias and/or diversity and serendipity.

\section*{Acknowledgements}
This work was partially supported by:
(i) SERICS (PE00000014) under the NRRP MUR program funded by the EU - NGEU;
(ii) MUR on D.M. 351/2022, PNRR Ricerca, CUP H23C22000440007, and (iii) MUR on D.M. 352/2022, PNRR Ricerca, CUP H23C22000550005. 




\bibliographystyle{plainnat}
\bibliography{ref}

\end{document}